\documentclass{aa}
\usepackage{natbib}
\usepackage{graphicx}
\usepackage{subcaption}
\usepackage{placeins}

\newcommand{\fgbm}{\textsl{Fermi}/GBM\xspace}
\newcommand {\swift} {\textsl{Swift}\xspace}
\newcommand {\fermi} {\textsl{Fermi}\xspace}
\newcommand {\nustar} {\textsl{NuSTAR}\xspace}

\newcommand {\integral} {INTEGRAL\xspace}
\newcommand {\chandra} {\textsl{Chandra}\xspace}
\newcommand {\xmm} {\textsl{XMM-Newton}\xspace}

\def \arcmin {\hbox{$^\prime$}}

\usepackage{txfonts}

\graphicspath{{./}{figures/}}

\begin{document}

\title{Extragalactic magnetar giant flare GRB~231115A: Insights from \textit{Fermi}/GBM observations}
\titlerunning{Extragalactic MGF GRB~231115A: Insights from \textit{Fermi}/GBM observations}
\author{Aaron C. Trigg\inst{1}
          \and Rachael Stewart\inst{2}
          \and Alex Van Kooten\inst{2}
          \and Eric Burns\inst{1}
          \and Matthew G. Baring\inst{3}
          \and Dmitry D. Frederiks\inst{4}
          \and Daniela Huppenkothen \inst{5}
          \and Brendan O'Connor\inst{6}
          \and Oliver J. Roberts\inst{7}
          \and Zorawar Wadiasingh\inst{8} \inst{9} \inst{10}
          \and George Younes \inst{8} \inst{11}
          \and Narayana Bhat\inst{12} \inst{13}
          \and Michael S. Briggs\inst{12} \inst{13}
          \and Malte Busmann\inst{14}
          \and Adam Goldstein\inst{7}
          \and Daniel Gruen\inst{14} \inst{15}
          \and Lei Hu\inst{6}
          \and Chryssa Kouveliotou \inst{16}
          \and Michela Negro \inst{1}
          \and Antonella Palmese\inst{6}
          \and Arno Riffeser\inst{14} \inst{17}
          \and Lorenzo Scotton\inst{12} \inst{13}
          \and Dmitry S. Svinkin\inst{4}
          \and Peter Veres\inst{12}
          \and Raphael Z{\"o}ller\inst{14} \inst{17}
          }

\institute{Department of Physics \& Astronomy, Louisiana State University, Baton Rouge, LA 70803, USA\\
            \email{atrigg2@lsu.edu}
        \and Department of Physics, The George Washington University, 725 21st Street NW, Washington, DC 20052, USA
        \and Department of Physics and Astronomy - MS 108, Rice University, 6100 Main Street, Houston, TX 77251-1892, USA
        \and Ioffe Institute, 26 Politekhnicheskaya, St. Petersburg, 194021, Russia
        \and SRON Netherlands Institute for Space Research, Niels Bohrweg 4, NL-2333 CA Leiden, the Netherlands
        \and McWilliams Center for Cosmology and Astrophysics, Department of Physics, Carnegie Mellon University, 5000 Forbes Avenue, Pittsburgh, PA 15213
        \and Science and Technology Institute, Universities Space and Research Association, 320 Sparkman Dr., Huntsville, AL 35805, USA.
        \and Astrophysics Science Division, NASA/GSFC, Greenbelt, MD 20771, USA
        \and Department of Astronomy, University of Maryland, College Park, MD 20742, USA
        \and Center for Research and Exploration in Space Science and Technology, NASA/GSFC, Greenbelt, MD 20771, USA
        \and CRESST, Center for Space Sciences and Technology, UMBC, Baltimore, MD 210250, USA
        \and Department of Space Science, University of Alabama in Huntsville, Huntsville, AL 35899, USA
        \and Center for Space Plasma and Aeronomic Research, University of Alabama in Huntsville, Huntsville, AL 35899, USA
        \and University Observatory, Faculty of Physics, Ludwig-Maximilians-Universität München, Scheinerstr. 1, 81679 Munich, Germany
        \and Excellence Cluster ORIGINS, Boltzmannstr. 2, 85748 Garching, Germany
        \and Department of Physics, George Washington University, Corcoran Hall, 725 21st Street NW, Washington, DC 20052, USA
        \and Max Planck Institute for Extraterrestrial Physics, Giessenbachstrasse, D-85748 Garching, Germany 
        }


    \abstract {Magnetar giant flares (MGFs) are the extremely short, energetic transients originating from highly magnetized neutron stars. When observed in nearby galaxies, these rare events are nearly indistinguishable from cosmological short gamma-ray bursts. We present the analysis of GRB\,231115A, a candidate extragalactic MGF observed by \fgbm and localized by \integral to the starburst galaxy M82. This burst exhibits distinctive temporal and spectral characteristics, including a short duration and a high peak energy, consistent with known MGFs. Time-resolved analysis reveals rapid spectral evolution and a clear correlation between luminosity and spectral hardness, providing robust evidence of relativistic outflows. Archival \textit{Chandra} data identified point sources within the GRB\,231115A localization consistent with the theoretical maximum persistent emission luminosity, though no definitive counterpart was found. Simulations indicate that any transient emission associated with GRB\,231115A would require energies exceeding those of typical magnetar bursts to be detectable by current instruments. While the tail of a MGF originating from outside of the Milky Way and its satellite galaxies has never been detected, analysis suggests that such emission could be observable at M82’s distance with instruments like \textit{Swift}/XRT or NICER, though no tail was identified for this event. These findings underscore the need for improved follow-up strategies and technological advancements to enhance MGF detection and characterization.}

\maketitle

\section{Introduction} \label{sec:intro} 

    Gamma-ray bursts (GRBs) are transient phenomena characterized by the emission of high-energy electromagnetic radiation in the 10\,keV to 100 GeV bands. These bursts can persist for durations ranging from a few milliseconds to several minutes \citep{kouveliotou2012gamma}. GRBs can potentially exhibit apparent luminosities that surpass those of typical supernovae by factors of hundreds and temporarily become the most luminous source of gamma-ray photons in the cosmos. GRBs are traditionally classified into two categories: short GRBs, which last less than two seconds, and long GRBs, which last more than two seconds. These two GRB populations differ not only in duration but also in their spectral properties \citep{Kouveliotou1993, AvK2020}. 
    
    Long GRBs make up the majority of current gamma-ray detections and are conventionally attributed to collapsars \citep{Woosley_annurev:/content/journals/10.1146/annurev.astro.43.072103.150558}. Short GRBs are attributed to two possible progenitors. The first, which makes up the vast majority of short GRBs, is the merger of compact objects, such as two neutron stars or a neutron star and a black hole~\citep{Eichler1989,Fong2015ApJ...815..102F}. These mergers may also produce a reasonable fraction of long GRBs \citep{Rastinejad-2022Natur.612..223R,Troja-2022Natur.612..228T,veres_extreme_2023}. The compact object merger theory was substantiated by the detection of a gravitational wave event on 17 August 2017, which coincided with a short GRB resulting from the coalescence of two neutron stars~\citep{Abbott2017a,Abbott2017b, Goldstein2017}. 
    
    The second source of short GRBs is the bright explosions from highly magnetized neutron stars \citep[i.e., magnetars;][]{DT1992,ThompsonDuncan1995} in nearby galaxies. Galactic magnetars have been observed emitting short, hard X-ray bursts and, on rare occasions, extremely energetic events known as magnetar giant flares \citep[MGFs;][]{kaspi2017}. MGFs are characterized by a brief, milliseconds-long spike in gamma-ray emission. This initial spike is far more energetic than those observed in typical magnetar burst spectra, with high isotropic-equivalent energies ($E_{\rm{iso}}\sim10^{44}-10^{46}\,\rm{erg}$). The bright, hard spike is followed by a much softer, weaker tail, which can last several minutes, modulated in flux and spectral hardness by the spin period of the neutron star from which it originates~\citep{hurley1999giant, palmer05:1806}. Only three such events have been detected and confirmed to date: two from Galactic sources, SGR 1900+14 \citep{hurley1999giant,Feroci1999} and SGR 1806-20 \citep{palmer05:1806,Frederiks2007AstL...33....1F}, and one from magnetar SGR 0526–66 in the Large Magellanic Cloud~\citep{mazets79Natur,Fenimore1996}. The initial peak of all three events was so bright that they saturated nearly all directly observing detectors. 
    
    At extragalactic distances beyond 3~Mpc, the rotationally modulated tail indicative of a MGF is too faint to be detected with current instruments due to their sensitivity limitations. Consequently, the absence of this tail would cause MGFs from nearby galaxies to resemble, and thus be misidentified as, cosmological short GRBs \citep{Mazets1982Ap&SS..84..173M,Eichler1989,Duncan2001AIPC..586..495D,hurley05:1806,palmer05:1806,Hurley2011}, presenting a challenge in identifying MGFs that originate outside the Milky Way. This misidentification may account for a portion of short GRB events, with one statistical estimate putting the number at $\sim$2\% \citep{2021Burns}. Therefore, localizing short GRBs that resemble MGFs to nearby star-forming galaxies is the most reliable method for differentiating MGFs from cosmological short GRBs.
    
    Currently, there are five short GRBs that, based on convincing evidence of their temporal and spectral characteristics, and with localizations by the Interplanetary Network \citep[IPN;][]{klebesdel1973ApJ...182L..85K} coinciding with nearby star-forming galaxies, have been identified as extragalactic MGF candidates. The first, GRB\,051103 \citep{ofek2006short,Frederiks_2007AstL...33...19F,Hurley_2010new}, was localized by IPN to an area on the outskirts of the galaxy M81 that shows evidence of tidal interactions with M82. Further multi-band observations \citep{ofek2006short,Frederiks_2007AstL...33...19F}, including searches for gravitational wave signals \citep{NAKAR2007166,Abadie_2012}, ruled out other progenitor classes as the origin of this burst, supporting the idea that GRB\,051103 was due to a MGF. Of the four remaining MGF candidates, two, GRB\,070201 and GRB\,070222, were found to have 2D spatial alignment with the nearby galaxies M31 and M83, respectively \citep{Mazets_2008,Ofek_2008,2021Burns}. The other two, GRB\,200415A and GRB\,180128A, were both associated with NGC\,253 \citep{svinkin2021bright,2021Natur.589..207R,Trigg2024A&A...687A.173T}, the first instance of two MGF candidates being localized to the same galaxy outside our own.

    On 15 November 2023, at 15:36:20.7 UT, the International Gamma-ray Astrophysics Laboratory (\integral) detected the short-burst GRB\,231115A \citep{Mereghetti2023,Minaev2024AstL...50....1M}. The \integral localization, at coordinates $\rm{R.A.}=149.0205^{\circ}$, $\rm{Dec.}=+69.6719^{\circ}$ (J2000, 2 arcmin 90$\%$ c.l. radius), promptly associated GRB\,231115A with the starburst galaxy M82 \citep{Mereghetti2023GCN.35037....1M}, identifying this burst as an extragalactic MGF candidate. Based on the \integral localization and using the method outlined in \cite{2021Burns}, the chance alignment as determined by a false alarm rate for GRB\,231115A is beyond $5\sigma$. This strong association with M82, which is greater than the association found for GRB\,200415A, makes GRB\,231115A the second extragalactic MGF candidate associated with a galaxy in the M81 Group.
    
    The Gamma-ray Burst Monitor \citep[GBM;][]{Dalessi2023GCN.35044....1D,meegan2009fermi} aboard the \fermi Gamma-ray Space Telescope, along with WIND/KONUS \citep{konus2023GCN.35062....1F,KONUS1995SSRv...71..265A}, Glowbug \citep{glow2023GCN.35045....1C,grove2020glowbug}, and the Hard X-ray Modulation Telescope (HXMT), known as Insight-HXMT \citep{insight2023GCN.35060....1X,INSIGHT2020SCPMA..6349502Z}, also detected GRB\,231115A. However, it was the prompt \integral galaxy association, identification, and public alert that enabled the unprecedented rapid follow-up observations of an extragalactic MGF candidate by the astrophysical community via NASA's General Coordinates Network\footnote{\url{https://gcn.nasa.gov/}} (GCN).

    As reported in \cite{Mereghetti2023}, follow-up observations were performed at X-ray wavelengths by the \swift X-Ray Telescope \citep[XRT;][]{burrows2000swift} and \textit{XMM-Newton}. Observations began at $T_{0}+9.0$~ks and $T_{0}+60$~ks for \textit{Swift}/XRT \citep{GCN35064} and \textit{XMM-Newton} \citep{GCN35175}, respectively. There was no confident detection of a new X-ray source within the burst localization to a depth of $\approx 4\times10^{-14}$ erg cm$^{-2}$ s$^{-1}$ in the $2-10$ keV energy range. Optical follow-up began as early as one hour after the gamma-ray trigger; in Table \ref{tab:follow-up_optical} we present a log of optical observations compiled from GCN Circulars \citep{GCN35052,GCN35056,GCN35055,GCN35057,GCN35078,GCN35067,GCN35077,GCN35051,GCN35092}, as reported in \cite{Mereghetti2023}.

    Further follow-up observations included searches for radio and gravitational wave signals. Despite its proximity ($\approx$~20 degrees) to the Canadian Hydrogen Intensity Mapping Experiment Fast Radio Burst Collaboration's (CHIME/FRB) field of view \citep{CHIME_1st_cat}, no radio emission was detected contemporaneously with the high-energy burst. Analysis using established methods constrained potential FRB-like radio emission from GRB\,231115A to $<260\,\rm{Jy}$ or $<720\,\rm{Jy \, ms}$, assuming a 10 ms pulse width, at the time of the \fgbm detection. This corresponds to a stringent upper limit on the radio spectral luminosity of $<3.8\times10^{30}\,\rm{erg \, s^{-1} \, Hz^{-1}}$ at a luminosity distance of 3.5~Mpc to M82 \citep{CHIME_FRB2023GCN.35070....1C}.

    Prompt limits from CHIME rule out FRB-like events contemporaneous with GRB\,231115A \citep{CHIME_FRB2023GCN.35070....1C}, both coincident with the \fgbm trigger (although the dispersive delay is unknown) and prior to GRB\,231115A. The lack of detection within 80 minutes prior to the \fgbm trigger, despite being directly overhead, provides us with additional constraints. These constraints limit the radio flux to $<0.5\,\rm{Jy}$ and the fluence to $<1.2\,\rm{Jy\,ms}$, assuming a 10 ms burst width, yielding a radio spectral luminosity limit of $<7.3\times10^{27}\,\rm{erg\,s^{-1}\,Hz^{-1}}$ \citep{CHIME_FRB2023GCN.35070....1C}.

    Gravitational wave observations were conducted at the Laser Interferometer Gravitational-Wave Observatory Hanford Observatory (H1) with an approximate average sensitive range of $\sim$150~Mpc for detecting binary neutron star mergers \citep{LVK_2023GCN.35049....1L}. Low-latency pipelines designed for identifying compact binary mergers were operational during this period. Despite this, as stated in \cite{Mereghetti2023}, no gravitational wave candidates were detected within a time window from -5~s to 1~s seconds around GRB\,231115A. Notably, the sensitivity of H1 extended to gravitational waves originating from the \integral sky position.
        
    Here we present the findings for GRB\,231115A. In Sect.~\ref{sec:prompt} we present various timing and spectral analyses in the gamma-ray regime of the prompt emission. In Sect.~\ref{sec:other} we discuss the various multiwavelength observations. Additionally, we comment on the search for persistent emission from the magnetar in M82 and calculate the distance at which we expect to see the tail. We discuss our interpretation of the physical mechanism of GRB\,231115A derived from our observations in Sect.~\ref{sec:discuss}. Finally, in Sect.~\ref{sec:conclusion} we provide our conclusions.

\section{Gamma-ray analysis} \label{sec:prompt}

    The \textit{Fermi}/GBM comprises 12 un-collimated thallium-activated sodium iodide (NaI) detectors and two bismuth germanate (BGO) detectors. The NaI detectors have an effective spectral range of approximately 8–900~keV, while the BGO detectors cover a range of approximately 250\,keV to 40\,MeV. These strategically placed detectors provide coverage of the unocculted sky across the full combined spectral range of 8~keV to 40~MeV \citep{meegan2009fermi}.

    In this study we analyzed the time-tagged event (TTE) data collected for GRB\,231115A. The TTE data, recorded with a temporal resolution of 2~$\mu$s, include the arrival time and energy channel (one of 128 channels) for each photon, with separate energy scales for the NaI and BGO detectors. The analysis was performed using the \textit{Fermi} Gamma-ray Data Tools \citep[GDT; v2.1.0;][]{GDT-Fermi}.
    
    Background estimation was conducted using the {\tt BackgroundFitter} module in GDT. This module, which allows the user to specify a polynomial fit, was used to fit the background data with a second-order polynomial over intervals preceding and following the burst, specifically from $T_{0}$-105~s to $T_{0}$-5~s and from $T_{0}$+5~s to $T_{0}$+105~s, where $T_{0}$ is the trigger time of the burst. It then interpolates the background data to time bins from $T_{0}$-105~s to $T_{0}$+105~s.
    
    \begin{figure}
        \resizebox{\hsize}{!}{\includegraphics{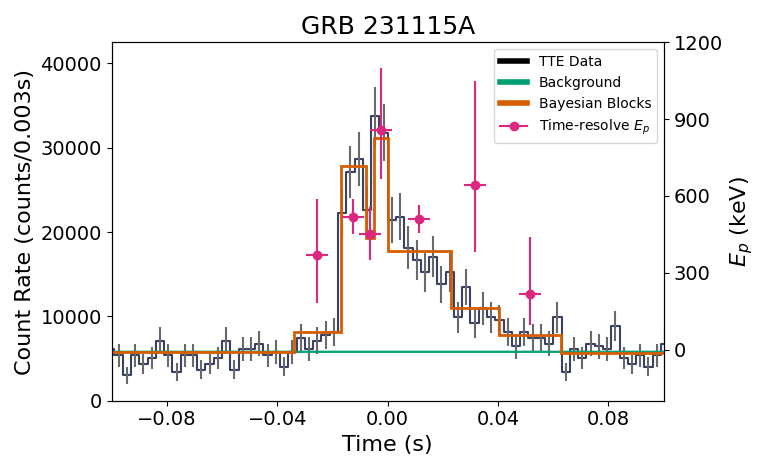}}
        \caption{\fgbm light curve of GRB\,231115A (black), binned at a temporal resolution of counts per 3~ms, with the background model (green). The brown line is the result of the BB analysis of the burst. The pink dots are the time-resolved peak energy values taken from Table\,\ref{tab:GBM_spec_BB}. At this finer temporal resolution we see a double-peaked structure within the initial burst signal.}
        \label{fig:GRB231115A_lc}
    \end{figure}

\subsection{Spectral analysis} \label{sec:spect}

     Using the more precise localization coordinates from \integral (see above), we generated responses for detectors observing that position within 60$^{\circ}$ of their boresight. We identify NaI detectors 3, 6, 7, 8, and 9, along with the BGO 1 detector, as those with the appropriate viewing angle. The light curve generated from these detectors' data, covering the energy range of 10--500 keV and overlaid with a Bayesian blocks \citep[BBs;][]{Scargle2013bayesian} analysis (in red), is shown in Fig.\,\ref{fig:GRB231115A_lc}. The light curve displays a multi-peak structure within the initial peak of GRB\,231115A. Similar variability has been seen in the other extragalactic MGF candidates GRB\,180128A \citep{Trigg2024A&A...687A.173T}, GRB\,200415A \citep{svinkin2021bright,2021Natur.589..207R}, GRB\,070201 \citep{Ofek_2008,Mazets_2008}, and GRB\,070222 \citep{2021Burns}.
    
    The $T_{90}$ duration, the interval between 5$\%$ and 95$\%$ of the cumulative fluence value, is $T_{90}=32\pm36$\,ms. The time between which 25$\%$ and 75$\%$ of the total fluence was accumulated is $T_{50}=16\pm23$\,ms. These values are determined by analyzing the background-subtracted light curve from the optimal detectors and fitting a Comptonized spectrum (see below) to track the fluence over time. Analyzing the event using BBs, we find a total burst duration ($T_{BB}$) of 97~ms. The rise time of the initial peak is $\sim2.7\,\rm{ms}$. As in \cite{2021Natur.589..207R}, we calculated the rise time of the pulse by fitting a pulse shape function and taking the elapsed time between the 10\%-90\% of the peak.
    
    We analyzed the GBM data over energies ranging from 8\,keV to 10\,MeV for GRB\,231115A and the KONUS data over energies ranging from 10\,keV to 1.2\,MeV for GRB\,051103\footnote{This analysis complements that of \cite{Frederiks_2007AstL...33...19F} and \cite{svinkin2021bright}, utilizing a different temporal binning}. We also reanalyzed the data from GRB\,180128A and GRB\,200415A using asymmetric errors as they better capture the uncertainties inherent in Poisson-distributed measurements, providing a more accurate representation than symmetric errors used in the previous analyses. These new analyses involve both time-integrated and time-resolved spectral fits.

    Both previously identified MGFs observed by \fgbm were analyzed using the RMfit software\footnote{The spectral analysis tool RMfit was initially designed for time-resolved analysis of BATSE GRB data. However, it has been adapted for GBM and other instruments supporting compatible FITS data formats. Information on the software can be accessed through the \textit{Fermi} Science Support Center at \url{https://fermi.gsfc.nasa.gov/ssc/data/p7rep/analysis/rmfit/}}, which utilizes an adapted forward-folding Levenberg–Marquardt algorithm for spectral fitting \citep{Goldstein_2013}. In this work, we instead employed the Nelder-Mead algorithm \citep{nelder1965simplex,gao2012implementing} for spectral fitting. While slower than Levenberg–Marquardt, Nelder-Mead provides a distinct advantage: if the true solution lies within the bounds of the initial simplex, convergence is guaranteed. This approach reduces the risk of becoming trapped in local minima. This is also a known limitation Levenberg–Marquardt.

    While our findings are generally consistent with those reported in \cite{2021Natur.589..207R} and \cite{Trigg2024A&A...687A.173T}, some differences are apparent, reflecting the improved fitting procedure. These updates provide a more accurate and well-constrained data characterization, particularly for GRB\,180128A, where the previous analysis encountered significant uncertainties.
    
    We fit the differential energy spectrum for GRB\,231115A to three different models. These models include a Band function \citep{1993ApJ...413..281B}, a power-law function (PL), and a Comptonized function \citep[COMPT;][]{gruber2014fermi}.  COMPT exhibits a power-law behavior characterized by an index $\alpha$, with an exponential cutoff at a characteristic energy $E_{\rm{p}}$ of the spectral peak of a $\nu F_{\nu}$ representation. 

    \begin{figure}
            \resizebox{\hsize}{!}{\includegraphics{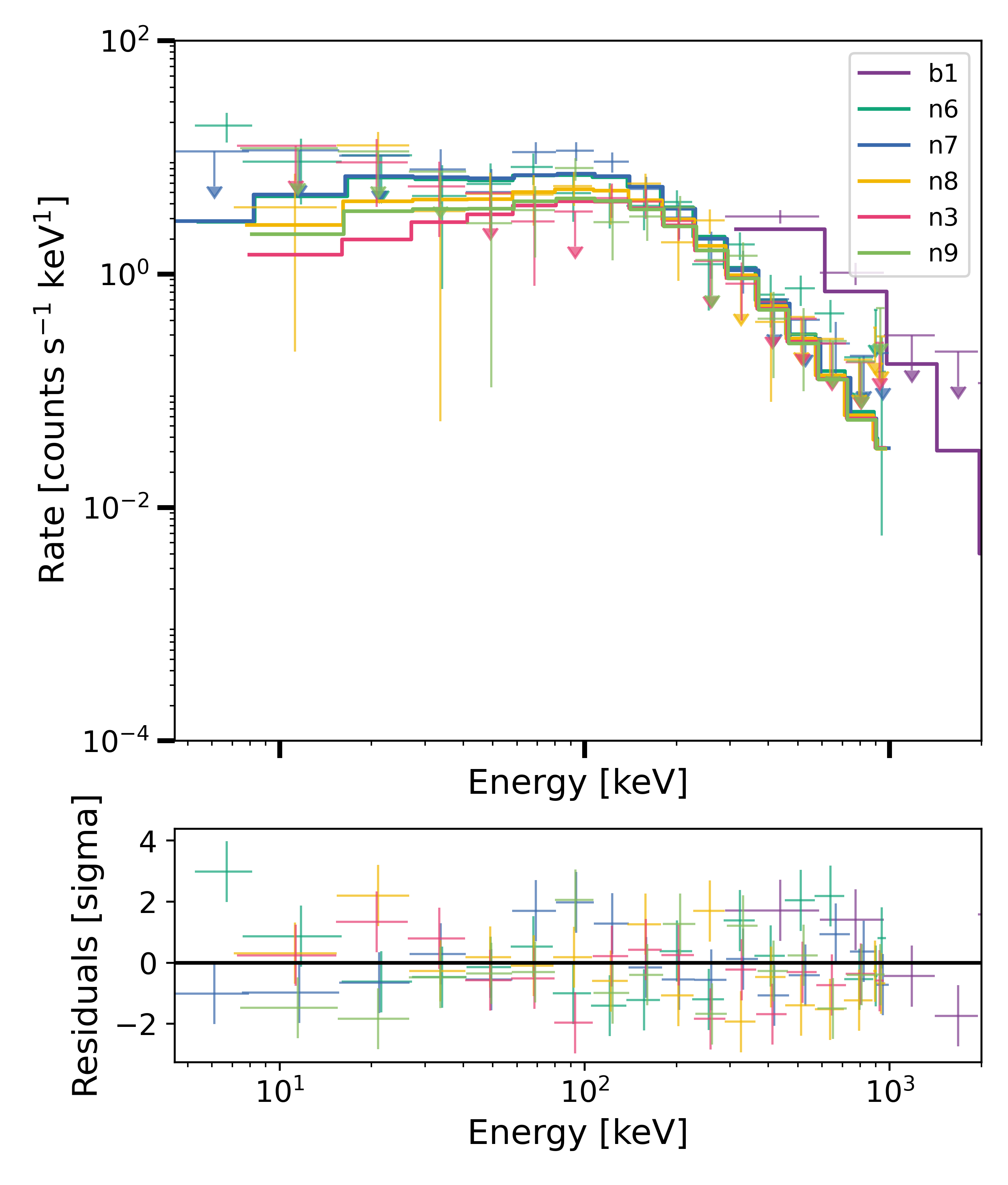}}
            \caption{Time-integrated spectral fit to a COMPT model for GRB\,231115A.}
            \label{fig:GRB231115A_t-i_spect}
    \end{figure}

    The three models were fit using the fit statistic pstat in GDT, which is a likelihood for Poisson data with assumed known background and is the same as the pstat statistic from the Xspec Statistics Appendix \citep{arnaud2011handbook}. Pstat was also used in the GBM spectral catalogs \cite{gruber2014fermi} and \cite{Poolakkil_2021}. We used the method in \cite{gruber2014fermi} to determine the best-fit model by comparing the $\Delta$\textit{C-}Stat (the difference in log-likelihood per degree of freedom) of the various models against the critical delta log-likelihood $\Delta$\textit{C-}Stat$_{crit}$ values listed in \citet[see our Table\,\ref{tab:pstat}]{gruber2014fermi}.

    We performed time-integrated fits for the significant emission durations as defined by the BB analysis for the bursts. The results of these time-integrated fits, reported in Table\,\ref{tab:GBM_spec_BB}, provide an overview of the spectral properties over the entire emission period. Based on the previously mentioned best-fit criteria, the time-integrated data are best-fit by the COMPT model (see Fig.\,\ref{fig:GRB231115A_t-i_spect}). Additionally, we conducted time-resolved fits using two approaches. The first method robustly determines the intervals by detecting and modeling changes in the gamma-ray emission rate, allowing the capture of detailed spectral evolution. The second approach uses intervals with equal durations to provide a consistent comparison across different phases of the bursts. The preferred model is PL for one interval in the time-resolved equal duration fit and four intervals in the time-resolved BB duration fit. The PL model preferred over the COMPT model is likely due to photon counts being too low to statistically prefer the curvature constraints of the COMPT model for those durations. However, we assumed that the COMPT model is the more accurate spectral form based on the time-integrated analysis. 

    The results of the time-resolved fits, presented at the 90\% confidence level, are listed in Tables \ref{tab:GBM_spec_BB} and \ref{tab:GBM_spec_fixed}, highlighting the temporal variation in spectral properties for GRB\,231115A and GRB\,051103. Figure \ref{fig:time-res_cont} illustrates the fitted spectra of GRB\,231115A corresponding to the data in Table \ref{tab:GBM_spec_BB}. The derived values of $\alpha$ and $E_{\rm{p}}$ for both bursts are consistent with those of previously identified MGFs and MGF candidates \citep{2021Burns,Trigg2024A&A...687A.173T}, with $\alpha$ typically ranging from approximately 0.0 to 1.0 and $E_{\rm{p}}$ spanning from $\sim$300 keV to several MeV. The analyses applied to the two previously identified extragalactic MGF candidates detected by \fgbm, GRB\,180128A and GRB\,200415A, are summarized in Tables \ref{tab:GBM_spec_other} and \ref{tab:GBM_spec_fixed}.
    
\begin{figure}
        \resizebox{\hsize}{!}{\includegraphics{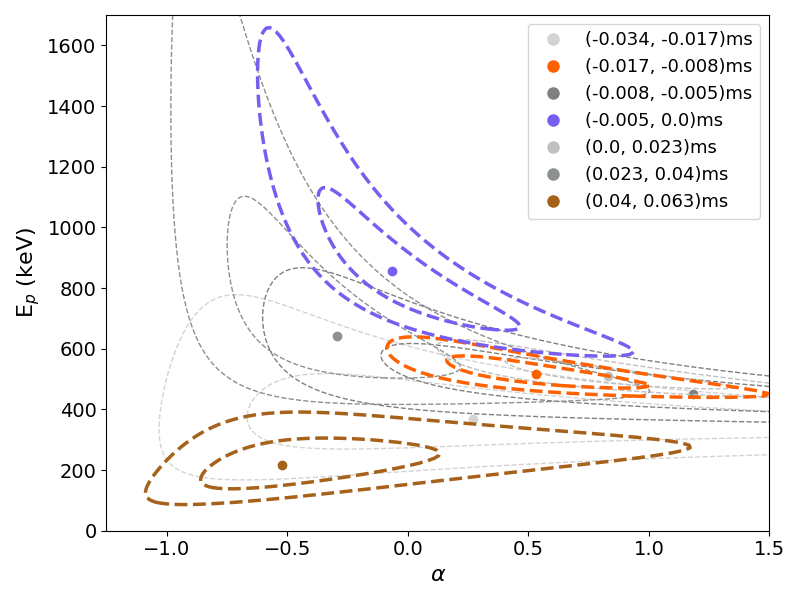}}
        \caption{Confidence contours for the time-resolved spectral analysis of GRB\,231115A, corresponding to the BB intervals in Table \ref{tab:GBM_spec_BB}. The contours represent the 50\% and 90\% confidence regions for the correlated parameters. Highlighted intervals illustrate spectral evolution within the burst.}
        \label{fig:GRB231115A_contour}
    \end{figure}

    The $E_{\rm{p}}$ values from the time-resolved analysis in Table\,\ref{tab:GBM_spec_BB} exhibit variations comparable to those observed in GRB\,200415A \citep{2021Natur.589..207R} and GRB\,180128A \citep{Trigg2024A&A...687A.173T}. The significance of this variation is illustrated by the highlighted intervals in Fig.\,\ref{fig:GRB231115A_contour}, which displays the 50\% and 90\% confidence intervals for the correlated parameters $\alpha$ and $E_{\rm{p}}$. The statistical significance of the observed variations was evaluated to determine whether they represent genuine spectral evolution or are consistent with random fluctuations. We conducted a rigorous statistical analysis to evaluate the significance of the deviations in $E_{\rm{p}}$ values across the time-resolved BB intervals in Table\,\ref{tab:GBM_spec_BB}. The time-integrated spectral fit, which provides a baseline of $E_{\rm{p}}=600\pm60$ (90\% confidence), was used as the reference for assessing the \textit{p}-values for each BB interval. While multiple individual intervals showed deviations from the baseline value, we calculated a joint \textit{p}-value to measure the collective significance of these deviations.

    The joint \textit{p}-value, obtained using Fisher’s combined probability test \citep{fisher1970statistical}, aggregates the \textit{p}-values of all the time-resolved intervals into a single statistic. This approach considers the overall pattern of deviations rather than treating each interval independently, making it particularly suited for evaluating trends in time-resolved spectral data. For GRB\,231115A, the resulting joint \textit{p}-value is 0.002, corresponding to $3\sigma$ significance. This low probability demonstrates that the observed evolution in $E_{\rm{p}}$ across the intervals is highly unlikely to result from random fluctuations, providing robust statistical confirmation of the spectral evolution.

{\renewcommand{\arraystretch}{1.3}
\begin{table*}
    \centering
    \caption{Time-resolved spectral analysis using BBs.}
    \label{tab:GBM_spec_BB}
    \begin{tabular}{ccccccc}
    \hline\hline
    Interval \# & Time & $E_{\mathrm{p}}$ & $\alpha$ & Energy Flux ($\cal{F}$) & $L_{\mathrm{iso}}$ &  $E_{\mathrm{iso}}$\\
     & (ms) & (keV) &  & ($\times10^{-6}$~ergs~s$^{-1}$~cm$^{-2}$) &($\times10^{45}$~erg$\cdot$~s$^{-1}$) &  ($\times10^{44}$~erg)\\ 
    \hline
    \multicolumn{6}{c}{GRB\,231115A}\\
    (1) & -34:-17 & 400 {\raisebox{0.5ex}{\tiny$\substack{+200 \\ -200}$}} & 0.3  {\raisebox{0.5ex}{\tiny$\substack{+3.8 \\ -1.4}$}} & 1.2  {\raisebox{0.5ex}{\tiny$\substack{+0.5 \\ -0.5}$}} & 1.8  {\raisebox{0.5ex}{\tiny$\substack{+0.7 \\ -0.3}$}} & 0.4  {\raisebox{0.5ex}{\tiny$\substack{+0.1 \\ -0.1}$}} \\ 
    (2) & -17:-8 & 520 {\raisebox{0.5ex}{\tiny$\substack{+70 \\ -70}$}} & 0.5  {\raisebox{0.5ex}{\tiny$\substack{+0.5 \\ -0.5}$}} & 20  {\raisebox{0.5ex}{\tiny$\substack{+2 \\ -2}$}} & 28.6  {\raisebox{0.5ex}{\tiny$\substack{+2.2 \\ -1.3}$}} & 2.75  {\raisebox{0.5ex}{\tiny$\substack{+0.19 \\ -0.13}$}} \\ 
    (3) & -8:-5 & 450 {\raisebox{0.5ex}{\tiny$\substack{+110 \\ -100}$}} & 1.2  {\raisebox{0.5ex}{\tiny$\substack{+1.5 \\ -1.7}$}} & 11  {\raisebox{0.5ex}{\tiny$\substack{+3 \\ -7}$}} & 16  {\raisebox{0.5ex}{\tiny$\substack{+4 \\ -2}$}} & 0.58  {\raisebox{0.5ex}{\tiny$\substack{+0.11 \\ -0.07}$}} \\ 
    (4) & -5:0 & 900 {\raisebox{0.5ex}{\tiny$\substack{+300 \\ -200}$}} & -0.1  {\raisebox{0.5ex}{\tiny$\substack{+0.4 \\ -0.4}$}} & 35  {\raisebox{0.5ex}{\tiny$\substack{+3 \\ -3}$}} & 51  {\raisebox{0.5ex}{\tiny$\substack{+3 \\ -2}$}} & 2.7  {\raisebox{0.5ex}{\tiny$\substack{+0.2 \\ -0.1}$}} \\ 
    (5) & 0:23 & 510 {\raisebox{0.5ex}{\tiny$\substack{+50 \\ -50}$}} & 0.8  {\raisebox{0.5ex}{\tiny$\substack{+0.5 \\ -0.5}$}} & 10.3  {\raisebox{0.5ex}{\tiny$\substack{+1.0 \\ -1.1}$}} & 15  {\raisebox{0.5ex}{\tiny$\substack{+1 \\ -1}$}} & 3.7  {\raisebox{0.5ex}{\tiny$\substack{+0.3 \\ -0.2}$}} \\ 
    (6) & 23:40 & 600 {\raisebox{0.5ex}{\tiny$\substack{+400 \\ -300}$}} & -0.3  {\raisebox{0.5ex}{\tiny$\substack{+0.7 \\ -0.6}$}} & 4.4  {\raisebox{0.5ex}{\tiny$\substack{+1.2 \\ -1.1}$}} & 6.4  {\raisebox{0.5ex}{\tiny$\substack{+1.3 \\ -0.9}$}} & 1.3  {\raisebox{0.5ex}{\tiny$\substack{+0.3 \\ -0.2}$}} \\ 
    (7) & 40:63 & 220 {\raisebox{0.5ex}{\tiny$\substack{+220 \\ -120}$}} & -0.5  {\raisebox{0.5ex}{\tiny$\substack{+1.5 \\ -0.6}$}} & 0.7  {\raisebox{0.5ex}{\tiny$\substack{+0.8 \\ -0.4}$}} & 1.0  {\raisebox{0.5ex}{\tiny$\substack{+0.9 \\ -0.6}$}} & 0.41  {\raisebox{0.5ex}{\tiny$\substack{+0.23 \\ -0.12}$}} \\ 

    \hline
     T$_{BB}$ duration (97): & -34:63 & 600 {\raisebox{0.5ex}{\tiny$\substack{+60 \\ -60}$}} & 0.14  {\raisebox{0.5ex}{\tiny$\substack{+0.24 \\ -0.25}$}} & 7.8  {\raisebox{0.5ex}{\tiny$\substack{+0.5 \\ -0.5}$}} & 11.4  {\raisebox{0.5ex}{\tiny$\substack{+0.6 \\ -0.4}$}} & 11.5  {\raisebox{0.5ex}{\tiny$\substack{+0.4 \\ -0.3}$}} \\ 

    \hline\hline
    \multicolumn{6}{c}{GRB\,051103}\\
    (1*) & -8:-2 & 2,100 {\raisebox{0.5ex}{\tiny$\substack{+7900 \\ -1300}$}} & -0.5  {\raisebox{0.5ex}{\tiny$\substack{+0.5 \\ -0.3}$}} & 200  {\raisebox{0.5ex}{\tiny$\substack{+450 \\ -110}$}} & 300  {\raisebox{0.5ex}{\tiny$\substack{+700 \\ -200}$}} & 17  {\raisebox{0.5ex}{\tiny$\substack{+39 \\ -9}$}} \\ 
    (2) & -2:4 & 1,700 {\raisebox{0.5ex}{\tiny$\substack{+1200 \\ -500}$}} & -0.2  {\raisebox{0.5ex}{\tiny$\substack{+0.2 \\ -0.2}$}} & 1,300  {\raisebox{0.5ex}{\tiny$\substack{+1000 \\ -400}$}} & 1,900  {\raisebox{0.5ex}{\tiny$\substack{+1400 \\ -600}$}} & 120  {\raisebox{0.5ex}{\tiny$\substack{+90 \\ -30}$}} \\ 
    (3**) & 4:10 & 10,000 {\raisebox{0.5ex}{\tiny$\substack{+0 \\ -7000}$}} & -0.36  {\raisebox{0.5ex}{\tiny$\substack{+0.20 \\ -0.08}$}} & 2,600  {\raisebox{0.5ex}{\tiny$\substack{+200 \\ -1500}$}} & 3,900  {\raisebox{0.5ex}{\tiny$\substack{+500 \\ -2000}$}} & 230  {\raisebox{0.5ex}{\tiny$\substack{+20 \\ -130}$}} \\ 
    (4*)  & 10:22 & 7,000 {\raisebox{0.5ex}{\tiny$\substack{+3000 \\ -5000}$}} & -0.39  {\raisebox{0.5ex}{\tiny$\substack{+0.28 \\ -0.11}$}} & 1,400  {\raisebox{0.5ex}{\tiny$\substack{+500 \\ -900}$}} & 2,000  {\raisebox{0.5ex}{\tiny$\substack{+800 \\ -1300}$}} & 240  {\raisebox{0.5ex}{\tiny$\substack{+80 \\ -180}$}} \\ 
    (5*) & 22:34 & 1,700 {\raisebox{0.5ex}{\tiny$\substack{+8000 \\ -800}$}} & 0.2  {\raisebox{0.5ex}{\tiny$\substack{+0.9 \\ -0.5}$}} & 280  {\raisebox{0.5ex}{\tiny$\substack{+960 \\ -140}$}} & 400  {\raisebox{0.5ex}{\tiny$\substack{+1400 \\ -300}$}} & 50  {\raisebox{0.5ex}{\tiny$\substack{+170 \\ -20}$}} \\ 
    (6) & 34:74 & 1,000 {\raisebox{0.5ex}{\tiny$\substack{+400 \\ -200}$}} & 0.4  {\raisebox{0.5ex}{\tiny$\substack{+0.5 \\ -0.3}$}} & 100  {\raisebox{0.5ex}{\tiny$\substack{+40 \\ -20}$}} & 150  {\raisebox{0.5ex}{\tiny$\substack{+50 \\ -30}$}} & 60  {\raisebox{0.5ex}{\tiny$\substack{+21 \\ -11}$}} \\ 
    (7) & 74:106 & 670 {\raisebox{0.5ex}{\tiny$\substack{+240 \\ -130}$}} & 0.1  {\raisebox{0.5ex}{\tiny$\substack{+0.4 \\ -0.3}$}} & 35  {\raisebox{0.5ex}{\tiny$\substack{+9 \\ -6}$}} & 52  {\raisebox{0.5ex}{\tiny$\substack{+13 \\ -9}$}} & 16  {\raisebox{0.5ex}{\tiny$\substack{4 \\ -3}$}} \\

    \hline
    T$_{BB}$ duration (114): & -8:106 & 1700 {\raisebox{0.5ex}{\tiny$\substack{+500 \\ -300}$}} &  -0.1 {\raisebox{0.5ex}{\tiny$\substack{+0.12 \\ -0.11}$}} &  250 {\raisebox{0.5ex}{\tiny$\substack{+80 \\ -40}$}} &  360 {\raisebox{0.5ex}{\tiny$\substack{+80 \\ -60}$}} &  490 {\raisebox{0.5ex}{\tiny$\substack{+110 \\ -60}$}}\\
    \hline\hline
    \end{tabular}
    \tablefoot{The BB interval, time-resolved fluence is from fitting the spectrum with a COMPT  over a combined \fgbm (NaI and BGO detectors) spectral range of 8 keV–40 MeV. For GRB\,051103 the fits were made using the WIND/KONUS three-channel spectra (20-1200 keV),
    which typically poorly constrain $E_{\rm{p}}$ outside this band. The $L_{\rm{iso}}$ and $E_{\rm{iso}}$ values were calculated over the standardized bolometric energy range of 1\,keV to 10\,MeV.
    For intervals marked with (**) the $E_{\rm{p}}$ is not constrained (i.e., $>$10 MeV). 
    For intervals marked with (*) the $E_{\rm{p}}$ positive error is not constrained.}
    \end{table*}}

\subsection{Highest-energy photons} \label{HEPhotons}

\begin{figure}
        \resizebox{\hsize}{!}{\includegraphics{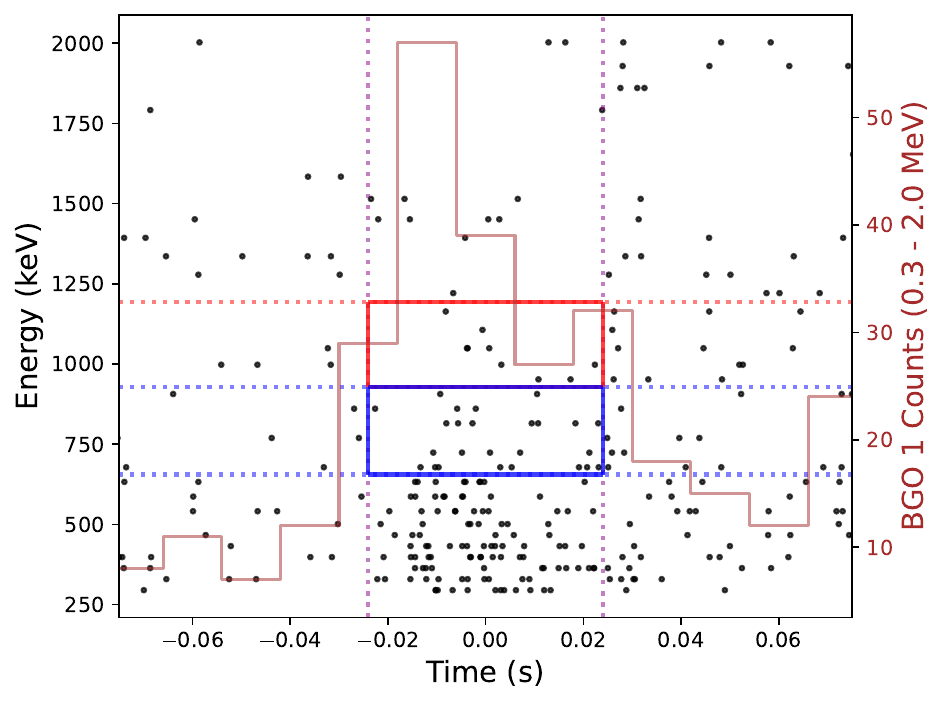}}
        \caption{Energetic photons from GRB\,231115A. Individual TTEs of GBM BGO detector 1 are indicated with black dots. The blue rectangle highlights energies from 655 keV to 930 keV, and the red rectangle energies from 930 keV to 1192 keV. We conclude that the highest photon energy associated with GRB\,231115A is $\sim$900\,keV.}
        \label{fig:HEphotonss}
    \end{figure}

        Figure\,\ref{fig:HEphotonss} shows the individual TTE counts in GBM BGO detector 1. GBM cannot determine with certainty whether any particular event is from GRB\,231115A, another gamma-ray source, or due to other background signals. We can assess whether a rate increase is statistically significant and thus likely associated with GRB\,231115A, using a Bayesian algorithm over Poisson data as a classic on-source:off-source method of source detection. This method is described in detail in \citet{2021Natur.589..207R} for MGF, GRB\,200415A. Over the 624 ms interval of our analysis, we find 110 TTE photons in a 576 ms background interval and 25 TTE photons in a 48 ms on-source window for energies ranging from 655 to 930 keV . The calculated probability for a source signal (GRB\,231115A) above the background is 0.999905 ($\sim4\,\sigma$), providing strong evidence that E$_{\rm{Max}}$ is $\sim930\,\rm{keV}$. For clarity, Fig.\,\ref{fig:HEphotonss} highlights a 140 ms window around T$_{0}$ to better display the on-source interval.

        We also considered a higher energy range, 930 to 1192 keV, for which there are 108 off-source TTE photons and nine on-source TTE photons (red box in Fig.~\ref{fig:HEphotonss}). The probability that the 9 on-source TTE photons represent an excess rate attributable to GRB\,231115A is 0.8545 ($\sim1.5\,\sigma)$, providing marginal evidence of the detection of E$_{\rm{Max}}\sim$1.1\,\rm{MeV} photons for GRB\,231115A by GBM, which is nearly one-third the energy of the highest-energy photons observed for GRB\,200415A~\citep{2021Natur.589..207R}, and somewhat similar to that reported by WIND solid-state telescope silicon detectors for the 27 December 2004 MGF initial pulse of SGR 1806-20~\citep{Boggs2007}. We therefore conclude that the highest photon energy associated with GRB\,231115A is $\sim900\,\rm{keV}$

\subsection{Minimum variability timescale} \label{mvt}

    The minimum variability timescale (MVT) is a measure of the shortest duration of significant fluctuation in the light curve of a GRB. Typically ranging from milliseconds to seconds, these timescales indicate rapid changes in the flux or intensity of a GRB. Given that MGFs can have bulk Lorentz factors ($\Gamma$) that are several orders of magnitude smaller than cosmological GRBs, we expect the MVTs to be much shorter for MGFs. The \cite{Golkhou_2015} method, based on Haar wavelets, yields an MVT of $T_{min}=1.1\,\pm\,0.7\,\rm{ms}$ for GRB\,231115A. An independent calculation identifies the MVT as the shortest binning timescale where the GRB signal is distinguishable from background fluctuations \citep{Bhat_2012}. This method produces a similar MVT value, $T_{min}=1.1\,\pm\,1.4\,\rm{ms}$. While this value is  unconstrained, the consistency between these two independently derived values strongly suggests that this is the true MVT for GRB\,231115A, enhancing the reliability of our measurement and the robustness of our method. The rise time value (Sect.\,\ref{sec:spect}) is broadly consistent with the MVTs $T_{rise}=2.7\,\pm\,1.1\,\rm{ms}$. These timescales correspond to an upper limit to the typical emission size, $cT \lesssim 3\times 10^{7} ~(T/{\rm ms}) ~{\rm cm}$. 

\subsection{Quasi-periodic oscillations} \label{QPOs}

    We searched the NaI and BGO data separately for quasi-periodic oscillations (QPOs). First, we generated light curves with a time resolution of $0.122$ ms, allowing us to search up to a frequency of 4096 Hz. We then created Leahy-normalized periodograms from those light curves and found that frequencies above 100 Hz are largely free of burst variability and thus consistent with white noise. We searched the frequency range from 100 to 4096 Hz using standard outlier detection techniques in the linearly and logarithmically binned periodograms and found no credible candidate detection at $p < 0.01$, corrected for the number of frequencies searched. 
    
    At frequencies below $100 \,\mathrm{Hz}$, we implemented the method described in \citet{huebner2022} and fit a model to the light curve containing an overall burst envelope parameterized as a skewed Gaussian as well as a damped random walk stochastic process. We compared that model to one that also includes a QPO parameterized as a stochastically driven damped harmonic oscillator. We compared models using the Bayes factor and define a strong candidate as one where $\log_{10}(\mathcal{B}) > 2$. We find a Bayes factor of $\log_{10}(\mathcal{B}_\mathrm{NaI}) = 0.24$ and $\log_{10}(\mathcal{B}_\mathrm{BGO}) = 0.49$ for the NaI and BGO, respectively, and conclude that there is no credible QPO candidate present in the data.

\section{Follow-up observations} \label{sec:other}

    Rapid follow-up observations provide a compelling case for GRB\,231115A as an extragalactic MGF. Below we detail our additional results and analyses, which support the conclusions in \cite{Mereghetti2023}.

\subsection{X-ray} \label{sec:x-ray}

    The arcminute localization uncertainty to GRB\,231115A offers a rare opportunity to search for the X-ray counterpart. Here, we briefly discuss the alternatives for such a search, focusing on the capabilities of current satellites.

\subsubsection{Persistent source} \label{sec:persist}

    Magnetars are persistent soft X-ray emitters, with emission potentially involving a heterogeneous, hot thermal surface modified by a highly magnetized atmosphere \citep[e.g.,][]{vigano13MNRAS}. Observations suggest a negative correlation between the surface temperature -- and consequently soft X-ray luminosity -- and the magnetar spin-down age within the population: younger magnetars are brighter and exhibit stronger bursting activity \citep[e.g.,][]{olausen14ApJS, vigano13MNRAS, enoto2019RPPh}. The brightest magnetars have persistent luminosities of $L_{\rm X, per} \gtrsim 10^{35}$~erg~s$^{-1}$. However, there is a maximum luminosity for magnetar soft thermal emission dictated by neutrino losses in the inner crust: $L_{\rm X, max} \lesssim 10^{36}$~erg~s$^{-1}$ \citep[e.g.,][]{pons12ApJ:mag}. This saturation restricts the detectability of such emissions from extragalactic magnetars at a distance of $\gtrsim3.5$~Mpc.
    
\begin{figure*}
        \centering
        \begin{minipage}[t]{0.44\textwidth}
            \includegraphics[width=\linewidth]{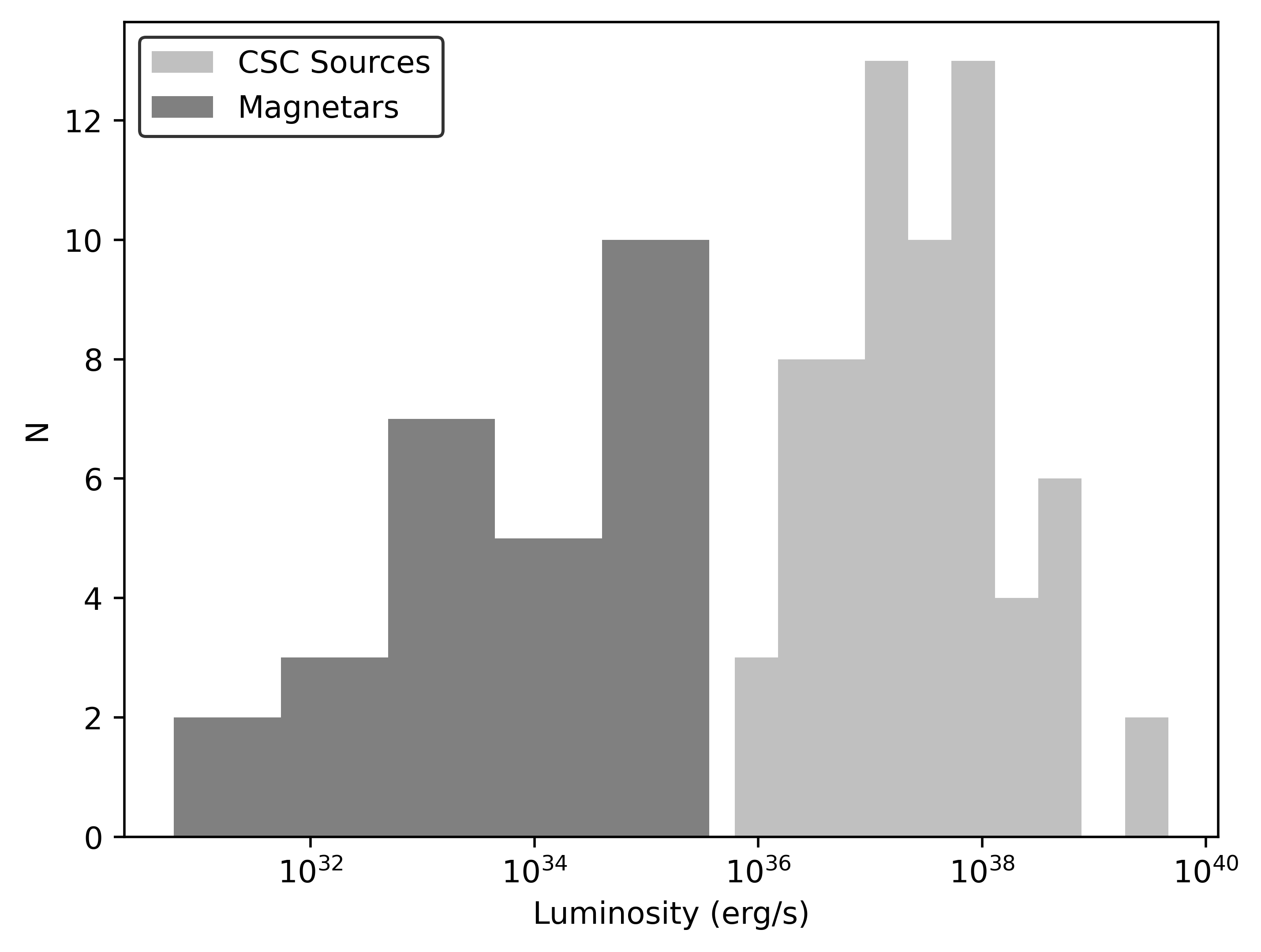}
        \caption{Luminosity distribution of M82 \chandra\ point sources within the \integral\ uncertainty region of GRB\,231115A (light gray histogram). They are detected in the full \chandra\ exposure of the galaxy, totaling $\approx1$~Ms. The dark gray histogram represents the luminosity distribution of known magnetars.}
        \label{psdist}
    \end{minipage}
    \hfill
    \begin{minipage}[t]{0.49\textwidth}
        \includegraphics[width=\linewidth]{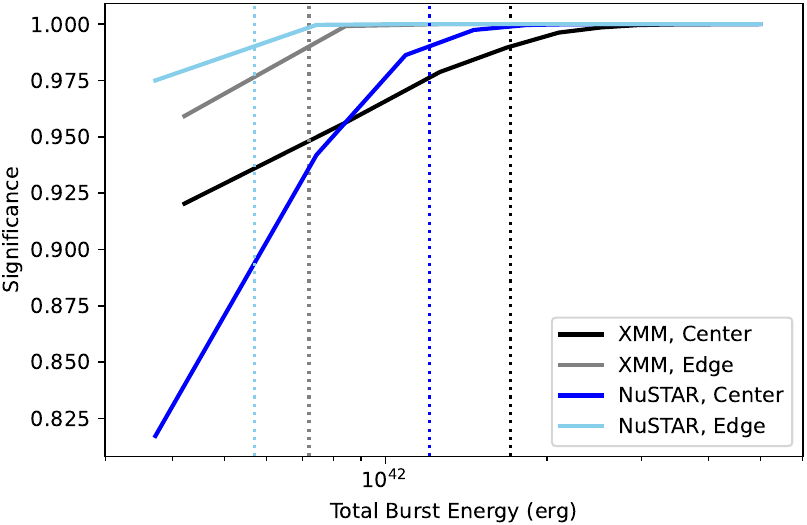}
        \caption{Detectability of short bursts with \xmm\ (black and gray lines) and \nustar\ (blue and cyan lines), if the putative magnetar is coincident with the center of the galaxy or offset by a few arcminutes, respectively. The vertical dotted lines correspond to burst energy for which a trial-corrected confidence level is $99\%$ for each case. See the main text for more details.}
        \label{burstdet}
    \end{minipage}
    \end{figure*}

     The galaxy M82 has been observed with \chandra throughout the mission, accumulating a total exposure of approximately 1~Ms. The luminosity distribution of the \chandra\ point sources within a $4\arcmin$ radius around the \integral burst position of GRB\,231115A, including foreground and background sources unrelated to M82, is shown in Fig.~\ref{psdist} (light gray). These data are taken from the \chandra\ Source Catalog\footnote{\url{https://cxc.cfa.harvard.edu/csc/}} (CSC). For comparison, the distribution of the known magnetar population is shown (dark gray), though it is heavily biased toward the brighter and most active magnetars. The faintest source detected with \chandra has a luminosity of $L_{\rm X}\approx10^{36}$~erg~s$^{-1}$, at the limit of the theoretical expectation for the maximum soft X-ray luminosity of a magnetar, though brighter by a factor of two than the brightest known magnetar (SGR 0526$-$66, $L_{\rm X}\approx5\times10^{35}$~erg~s$^{-1}$; \citealt{kulkarni03ApJ, park12ApJ}). While a few of these faint sources might be good magnetar candidates, without enough counts for an adequate spectral and, most importantly, temporal analysis, it is difficult to determine their true nature. Given the high star formation rate in M82, other possible source types include young pulsars, accreting pulsars, high-mass X-ray binaries, and young supernova remnants.

\subsubsection{Bursting magnetar} \label{sec:bursting}
    
    Given that all three confirmed MGFs were accompanied by short bursts occurring within hours to days following the event \citep{mazets79Natur, woods99ApJ, Frederiks2007AstL...33....1F}, we devised a methodology to blind-search for short bursts in imaging data such as \textit{XMM-Newton} European Photon Imaging Camera (EPIC-PN) and \textit{NuSTAR} focal plane modules (FPMs) as follows. We produced cleaned images in the 2-10 keV for PN and 3-70 keV ranges for FPMs, centered at the \integral localization $\pm$4\arcmin. We searched consecutive one-second frames (with a 0.2~s overlap) within valid good-time intervals. For this search, we used the SAS tool \texttt{edetect\_chain}\footnote{\url{https://xmm-tools.cosmos.esa.int/external/sas/current/doc/edetect_chain/}} for PN and the \texttt{DETECT} tool, part of the XIMAGE software for FPMs \citep{perri2013nustar}. Once these tools flag candidate sources, we compared their excess counts, $n$, to the local background rate, $\lambda$, determined from the full observation, and calculated the probability that the counts occur randomly, $P_{\rm i} = \lambda^{n_{\rm i}} e^{-\lambda} / n_{\rm i}!$, where index $i$ is for each frame searched. Frames with candidate sources having probability $P_{\rm i}<0.01/N$, where N is the number of frames searched (approximately equal to the live-time in each observation), are considered valid sources at the 99\% confidence level (trial corrected). We find no source that meets our criteria for a confident detection in either \textit{XMM-Newton} or \textit{NuSTAR}.
    
    We established the upper limit on the detectability of short bursts in these observations through simple simulations. We assumed that the magnetar short bursts' light curves have a fast-rise exponential-decay shape, with rise and decay times of $0.5\pm0.3$ and $0.7\pm0.3$~seconds, respectively \citep[e.g.,][]{younes20ApJ1935}. We assumed the burst spectra to follow a two-blackbody model with temperatures of 5 and 11~keV \citep[e.g.,][]{lin20ApJ}. Finally, we varied the 1-100~keV burst energy between $10^{41}$~erg and $10^{43}$~erg logarithmically, in 20 steps. For each simulated burst, the estimated number of counts and their corresponding times and energies are injected into the actual \textit{XMM-Newton} (or \textit{NuSTAR}) event file, and our search methodology is repeated. As two extreme cases, we injected the simulated "source" at the edge of the \integral error circle, away from point sources, and at the galaxy's bright center. 
    
    We find that the short burst fluence detection thresholds, according to our criterion, are $\gtrsim4\times10^{-10}$~erg~cm$^{-2}$ and $\gtrsim10^{-9}$~erg~cm$^{-2}$ for the edge and center cases, respectively. This translates to 1-100~keV energies of approximately $6\times10^{41}$~erg~s$^{-1}$ and $1.5\times10^{42}$~erg~s$^{-1}$, respectively (Fig.~\ref{burstdet}). These figures are similar for \xmm\ and \nustar, and are at the high end of typical short burst energetics, approaching the energies of intermediate flares.
    
    Due to the steep $\log N$-$\log S$ distribution of bursts, our non-detection with PN and FPMs is not surprising. Better prospects can be achieved with the High Energy X-ray Probe \citep[HEX-P;][]{Madsen2019HEX}, yet still at the level of the brightest short magnetar bursts \citep{alford24FrASS}.

\subsubsection{MGF tail detection}\label{sec:tail}
    
    Currently, the combination of \textit{Swift}'s Burst Alert Telescope \citep[BAT;][]{Barthelmy05} and autonomous re-pointing to observe with the \swift/XRT is the most promising means by which to capture a MGF tail in time for a possible detection. Unfortunately, for the existing catalog of extragalactic MGF candidates, the source was outside the BAT coded field of view, preventing a possible BAT trigger and the seconds-scale automatic re-pointing of \textit{Swift}. Another possible avenue is NICER's new capability for automatic re-pointing via a MAXI trigger \citep[OHMAN;][]{gendreau23HEAD}. However, this is limited by the low rate of short-GRB detection with MAXI, given its limited energy range $2-30$~keV, and NICER visibility, which is complex due to the structure of the International Space Station. 
    
\begin{figure*}
    \begin{center}
    \includegraphics[width=17cm]{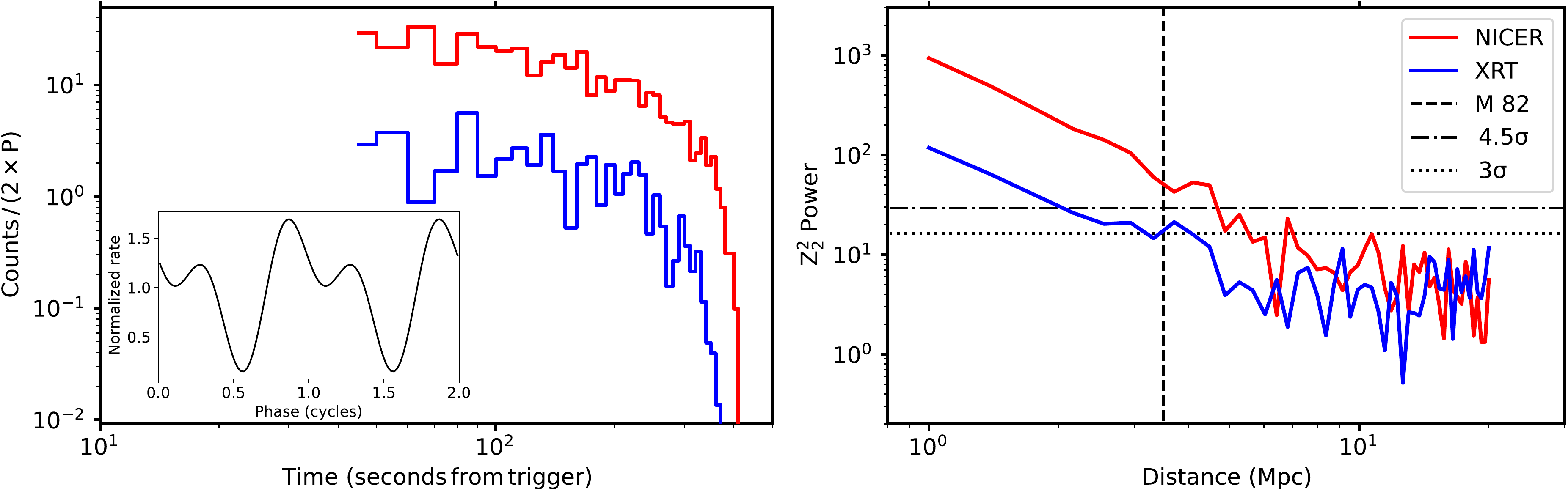}
    \caption{Re-pointing simulation results. \textit{Left}: Simulated MGF tail light curve with XRT (blue) and NICER (red) assuming the properties as observed in the SGR 1806$-$20 MGF tail, and scaling it to the M82 distance of 3.5~Mpc. The inset shows the modulation that is embedded in the light curve. \textit{Right}:  $Z_2^2$ power for pulsation detection in XRT and NICER as a function of MGF distance. The dot-dashed and the dotted lines show the 4.5$\sigma$ and 3$\sigma$ detection significance (single-trial), respectively. The vertical dashed line is the M82 distance of 3.5~Mpc. See the main text for more details.}
    \label{simTail}
    \end{center}
    \end{figure*}

    To assess the detectability of MGF tails with \textit{Swift} and a NICER-like instrument, we simulated the 2004 MGF tail from SGR~1806$-$20 as presented in \citet[see also \citealt{palmer05:1806}]{hurley05:1806}. The spectrum is modeled as a blackbody with a constant temperature of about 8 keV, folded through the response matrices and effective area curves of \textit{Swift}/XRT and NICER. The decay follows the evaporating fireball model $L_{\rm x}(t)=L_0[1-(t/t_{\rm evap})]^{a/(1-a)}$ with $L_0=10^{42}$~erg~s$^{-1}$ in the 20-100 keV energy range measured at $t\approx40$~s after the initial spike, $t_{\rm evap}=382$~s, and $a=0.6$ \citep{hurley05:1806}. Finally, we modulated the decaying tail at a fiducial spin period of $P=5$~s with a pulse shape following a Fourier series with two harmonics (having approximately equal power; see the inset of Fig.~\ref{simTail}) and an rms pulsed fraction of $45\%$. An example light curve for \textit{Swift} and NICER at the M82 distance of 3.5~Mpc is shown in the left panel of Fig.~\ref{simTail}.

    The detectability of the tail at extragalactic distances will largely depend on the re-pointing time and the background rate of the instrument \citep{negro2024rolemagnetartransientactivity}. The latter is $\approx5\times10^{-3}$~counts~s$^{-1}$ and $\approx5\times10^{-1}$~counts~s$^{-1}$ for XRT and NICER in the energy range 0.5-10 keV, respectively. For each light curve, we derived the cumulative counts in the tail from a start time ranging from 40 to 300 seconds and compared them to the cumulative background counts in the same time interval. We find that a $3\sigma$ detection (based on a Poisson probability density function for XRT and a Gaussian probability density function for NICER) can be achieved for re-pointing times of $\lesssim250$~seconds for both instruments. 
    
    However, pulsation detection, shown in the right panel of Fig.~\ref{simTail} in the form of the $Z^2_2$ power, is more prominent in NICER compared to XRT. Varying the distance from 1~Mpc to 20~Mpc reveals that a NICER-like instrument can detect the pulsations in the tail up to a distance of $\sim$6~Mpc, while an XRT-like instrument can do so up to 4~Mpc (Fig.~\ref{simTail}, right panel). One caveat is that XRT needs to operate in windowed-timing mode to detect pulsations $<5$~seconds due to its limited timing resolution of 2.5~seconds when operating in photon-counting mode.
        
    Identifying the progenitor magnetar of an extragalactic MGF within a few megaparsecs hinges on detecting the X-ray tail, which requires a rapid, autonomous response from instruments such as XRT. XRT provides arcsecond localization, which is crucial for studying the magnetar environment — a challenge in the Milky Way due to higher column densities and uncertain distances to Galactic magnetars. The X-ray follow-up of triggers from other satellites is also critical; for example, XRT or NICER could have succeeded in an automated follow-up of an \integral\ event. While less probable, a detection by NICER through the OHMAN program could potentially represent the first observation of a pulse period from an extragalactic magnetar. With incremental improvements to currently available technology, the prospects for such detections could significantly improve, for example by utilizing detectors like the Low Energy Modular Array and the High Energy Modular Array on the proposed probe-class mission STROBE-X \citep[Spectroscopic Time-Resolving Observatory for Broadband Energy X-rays;][]{StrobeX2019BAAS...51g.231R}.

\subsection{Optical} \label{sec:optic}

    Here, we present our optical follow-up with the 2.1m Fraunhofer telescope \citep{2014SPIE.9145E..2DH} at Wendelstein Observatory, Germany.
    
    Observations with the Three Channel Camera \citep[3KK;][]{2016SPIE.9908E..44L} at Wendelstein Observatory began on 16 November 2023 at 00:39 UT in the $riJH$ filters for two hours. 
    Additional observations were performed on 22 November 2023 at 00:16 UT with the 3KK camera simultaneously in the $riJ$ filters for one hour, and the Wendelstein Wide Field Imager \citep[WWFI;][]{2014ExA....38..213K} in the $g$ filter for one hour. 
    
    Data reduction was performed using a custom pipeline developed at the University Observatory Munich, for both the WWFI and 3KK cameras \citep{2002A&A...381.1095G,2020PhDT........30K,2020ApJS..247...43K}. The pipeline corrects for bias, dark, and flat-field properties and detector artifacts. It uses \texttt{SCAMP} \citep{Bertin2006} to compute the astrometric solution with respect to the \textit{Gaia} Early\ Data Release 3 catalog \citep{2021A&A...649A...1G}, and \texttt{SWarp} \citep{2002ASPC..281..228B,2010ascl.soft10068B} to co-add images. 
    Photometric zero points were calibrated using nearby stars in the Pan-STARRS-3 Pi (PV3; \citealt{2013ApJS..205...20M}) source catalog and Two Micron All Sky Survey (2MASS; \citealt{Skrutskie2006}) catalog.

    Difference imaging was conducted using the Saccadic fast Fourier transform algorithm\footnote{\url{https://github.com/thomasvrussell/sfft}} \citep{SFFT}. Archival templates from the WWFI, obtained in 2014 in $g$ and $r$ bands, were used as references. For other filters ($i$ and $J$) we performed difference imaging between images acquired on 16 November 2023 and 22 November 2023. Our analysis did not reveal and confidently detect rapidly varying transients between epochs. The $5\sigma$ upper limits for each filter are reported in Table \ref{tab:follow-up_optical}.

\section{Discussion} \label{sec:discuss}

    \begin{figure}
        \resizebox{\hsize}{!}{\includegraphics{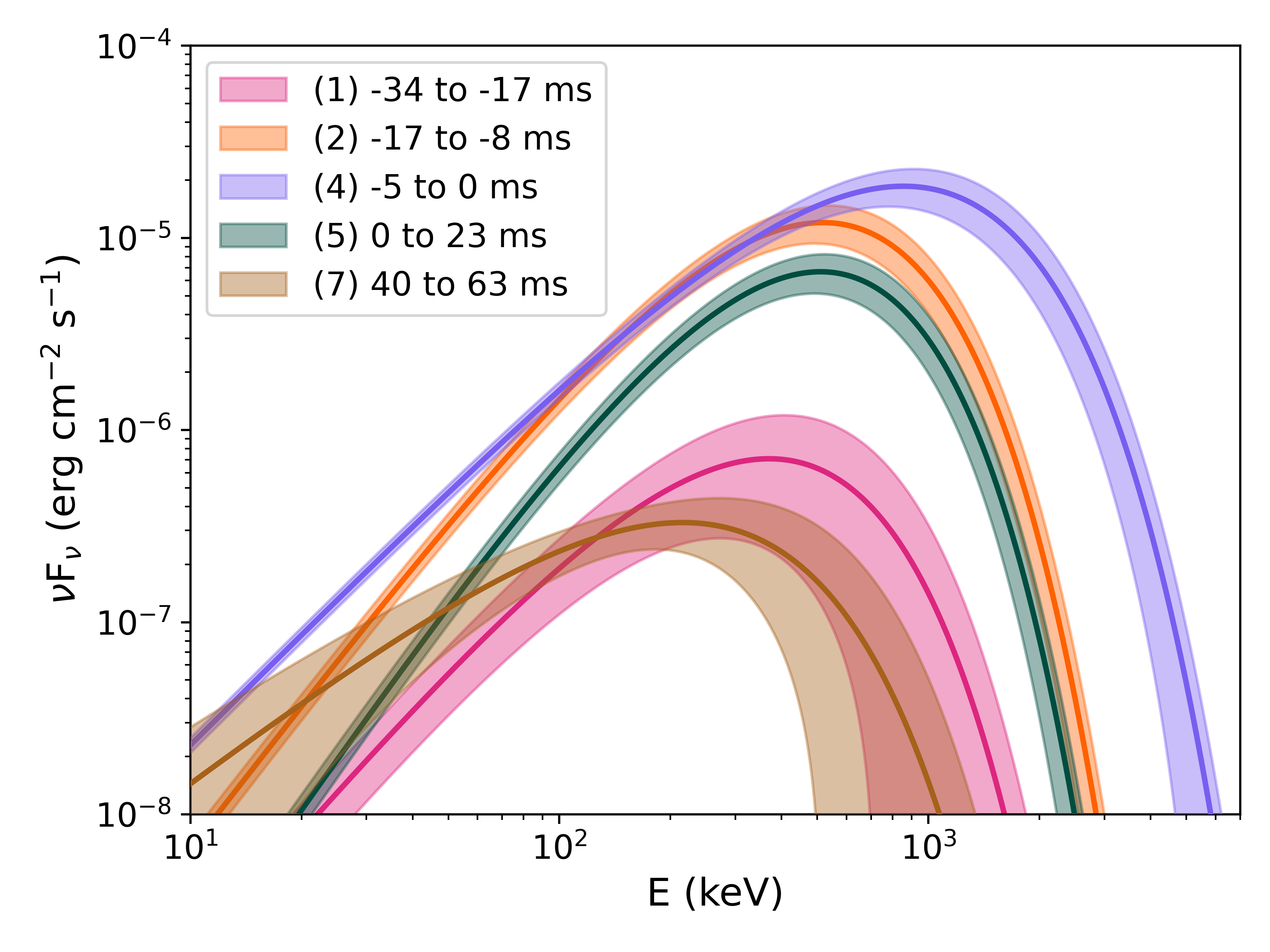}}
        \caption{Spectra of GRB\,231115A over four BB time intervals. The intervals show the onset of the burst (1), peak 1 (2), peak 2 (4), and the extended emission after the peaks (5). Intervals (3), (6), and (7) are omitted for clarity but are consistent with the trend displayed. The shaded area indicates the 90\% confidence regions}
        \label{fig:GRB231115A_nuFnu}
    \end{figure}
    
    While the sky association of GRB\,231115A with the nearby galaxy M82 is strongly suggestive of a non-cosmological GRB origin for this transient, its spectral evolutionary character provides additional localization-independent arguments for it being a MGF. This is predicated on similarities of its spectral evolution with those of the two MGFs from the Sculptor galaxy, NGC\,253, namely GRB\,180128A \citep{Trigg2024A&A...687A.173T} and GRB\,200415A \citep{2021Natur.589..207R}. The character of this evolution is naturally expected for an intense radiation beam from a rotating magnetar, a model that was highlighted in \cite{2021Natur.589..207R}. In the absence of an oscillatory tail (discussed in Sect.\,\ref{sec:tail}), the smoking gun signature of a MGF, the initial spike spectral evolution is the main information that can yield insights into the source of these transients.
    
    The rotating, relativistic ``lighthouse'' picture presumes that the plasma that powers the MGF is blasted off the magnetar surface near the magnetic poles and escapes as a wind to high altitudes and beyond the magnetosphere \citep[e.g.,][]{ThompsonDuncan1995}. This ejection could be triggered by the buildup of magnetic stresses in the crust that eventually force it to crack and release copious amounts of plasma and energy. The highly super-Eddington environment and enormous energy density forces the plasma to flow out at relativistic speeds. This expectation is commensurate with lower limits to the bulk Lorentz factor ($\Gamma$) of the outflow that is obtained from the argument that the emission region is transparent to $\gamma\gamma\to e^+e^-$ pair creation for all photons up to the maximum energy, $E_{\rm max}$ observed. This transparency is aided by Doppler beaming of the radiation \citep{Krolik-1991-ApJ,Baring-1993-ApJ}, and $\Gamma>E_{\rm max}/511\,$keV is the most conservative bound possible. Transparency of the emission to two-photon pair creation up to E$_{max}\sim$\,900\,keV (nearly double the threshold for pair creation) guarantees that the bulk Lorentz factor of the emitting plasma is at least around $\Gamma \sim 2$ relative to the observer, which is less than $\Gamma>6$ that \cite{2021Natur.589..207R} obtained for GRB\,200415A, which had $E_{\rm max}\sim 3\,$MeV. In contrast, GRB\,180128A was somewhat fainter with no photons above 511 keV in energy and above the background level, so $\Gamma>1$ is thereby unconstrained.
    
    The deduced relativistic motion of the plasma automatically implies that the brightness of the flare will vary rapidly and be correlated with its spectral evolution in a fairly well-defined manner dictated by standard special relativistic transformations. As the radiation--plasma beam sweeps across the line of sight to Earth, the luminosity should first intensify and the spectrum should harden to higher energies, and then both should recede (soften) as the beam rotates away from the observer \citep{2021Natur.589..207R}. This is the evolutionary sequence seen in both MGF candidates, GRBs\,180128A and GRB\,200415A. 
    
    To illustrate that GRB\,231115A exhibits the evolutionary sequence discussed in above, and verified statistically in Sect.\,\ref{sec:spect}, we plotted four of the BB intervals (the $E_{\rm{p}}$ values for all of the BB intervals are plotted in on the light curve in Fig.\,\ref{fig:GRB231115A_lc}). The results, shown in Fig.\,\ref{fig:GRB231115A_nuFnu}, showcase the spectral evolution from the onset of the burst through the two peaks and into the brief emission that follows. For clarity, the third, sixth, and seventh BB intervals have been omitted from Fig.\,\ref{fig:GRB231115A_nuFnu}, although they align with the overall trend seen in the included intervals. This spectral evolution, characteristic of a MGF, is clearly apparent for GRB\,231115A. Such ``envelope'' behavior is primarily a consequence of the Doppler beaming and boosting from the radiating plasma. This behavior does not naturally arise for classical GRBs born from collapsars or neutron star-neutron star mergers, and so one can conclude that GRB\,231115A is likely the initial spike of a MGF.

     \begin{figure}
            \resizebox{\hsize}{!}{\includegraphics{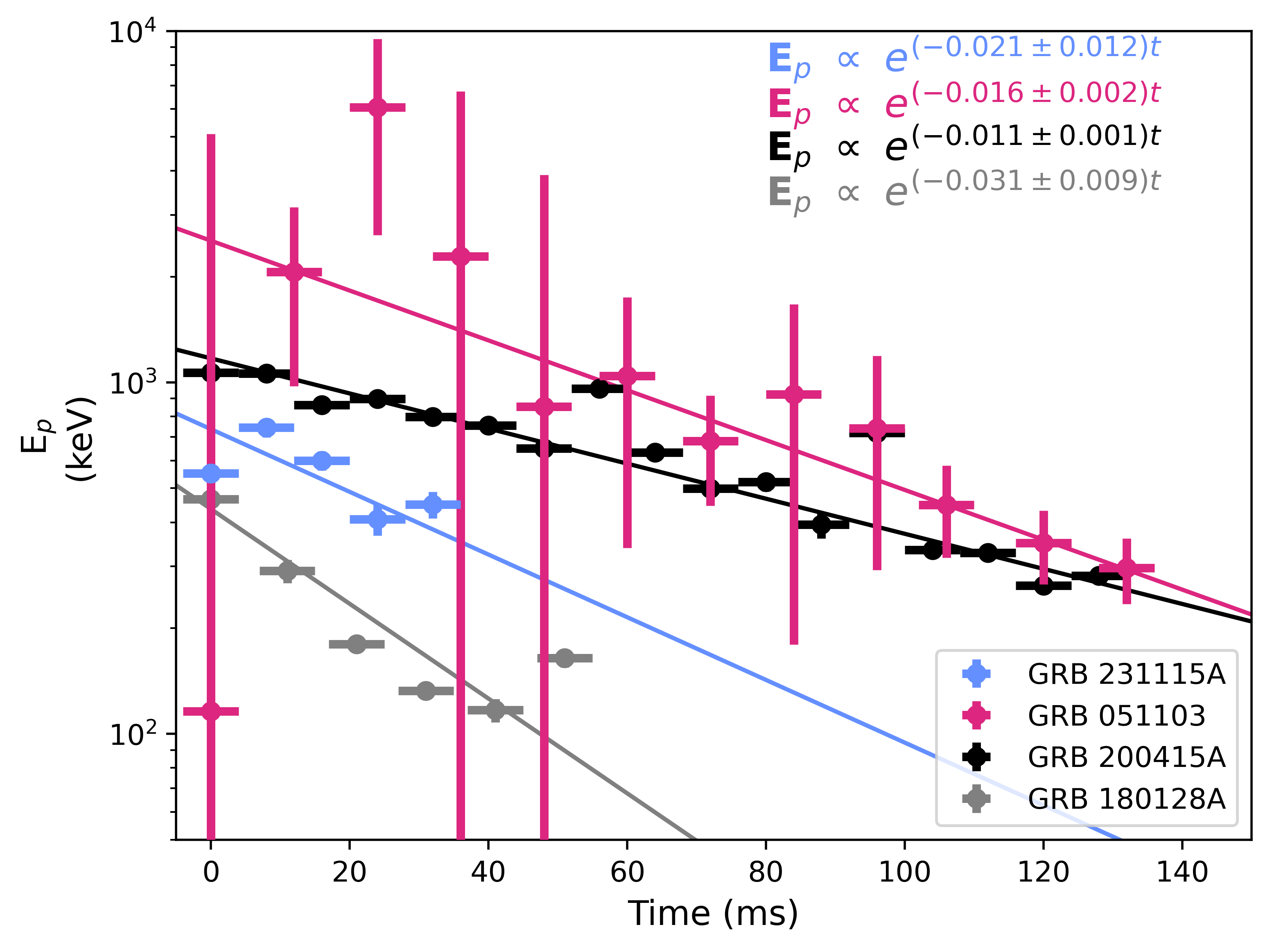}}
            \caption{Comptonized spectrum peak energy ($E_{\mathrm{p}}$) as a function of time using a temporal binning of 8\,ms and values from Table\,\ref{tab:GBM_spec_fixed}. All fit errors and error bars are at the 90\% confidence level. The zero-time reflects the GBM event start time of each detection.}
            \label{fig:GBM_Ep_time}
    \end{figure}
    
    The $\Delta t \sim 40\,$ms timescale for the overall evolution in Fig.~\ref{fig:GBM_Ep_time} can be used to provide a lower bound to the magnetar's rotation period. This time corresponds to a stellar rotation through an angle $\Delta \theta \sim 2\pi \Delta t/P$ for a rotation period of $P$. For a rotating beam of relativistic plasma, the peak flux of the emission light curve will be approximately confined to the Doppler cone of opening angle $1/\Gamma $, as long as the angular extent of the wind's collimation is not large. Combining these, one estimates that the putative neutron star's spin period should be bounded by $P\sim2\pi \Gamma \, \Delta t \gtrsim 500\,$ms for $\Gamma>2$ for GRB\,231115A. In practice, the bulk motion is expected to be somewhat or significantly faster than the pair creation transparency bound suggests. This is mainly due to the plasma dynamics associated with the large amount of energy deposited into the inner magnetosphere over short timescales. It is notable that the detection of delayed GeV-band emission in association with the GRB\,200415A giant flare suggested that $\Gamma\sim100$ was likely \citep{LATMGF}. If $\Gamma$ lies in the $3-50$ range for GRB\,231115A, then the deduced period $P$ would be commensurate with those of Galactic magnetars.

    \begin{figure*}
        \begin{minipage}{.5\textwidth}
            \begin{center}
            \includegraphics[width=\linewidth]{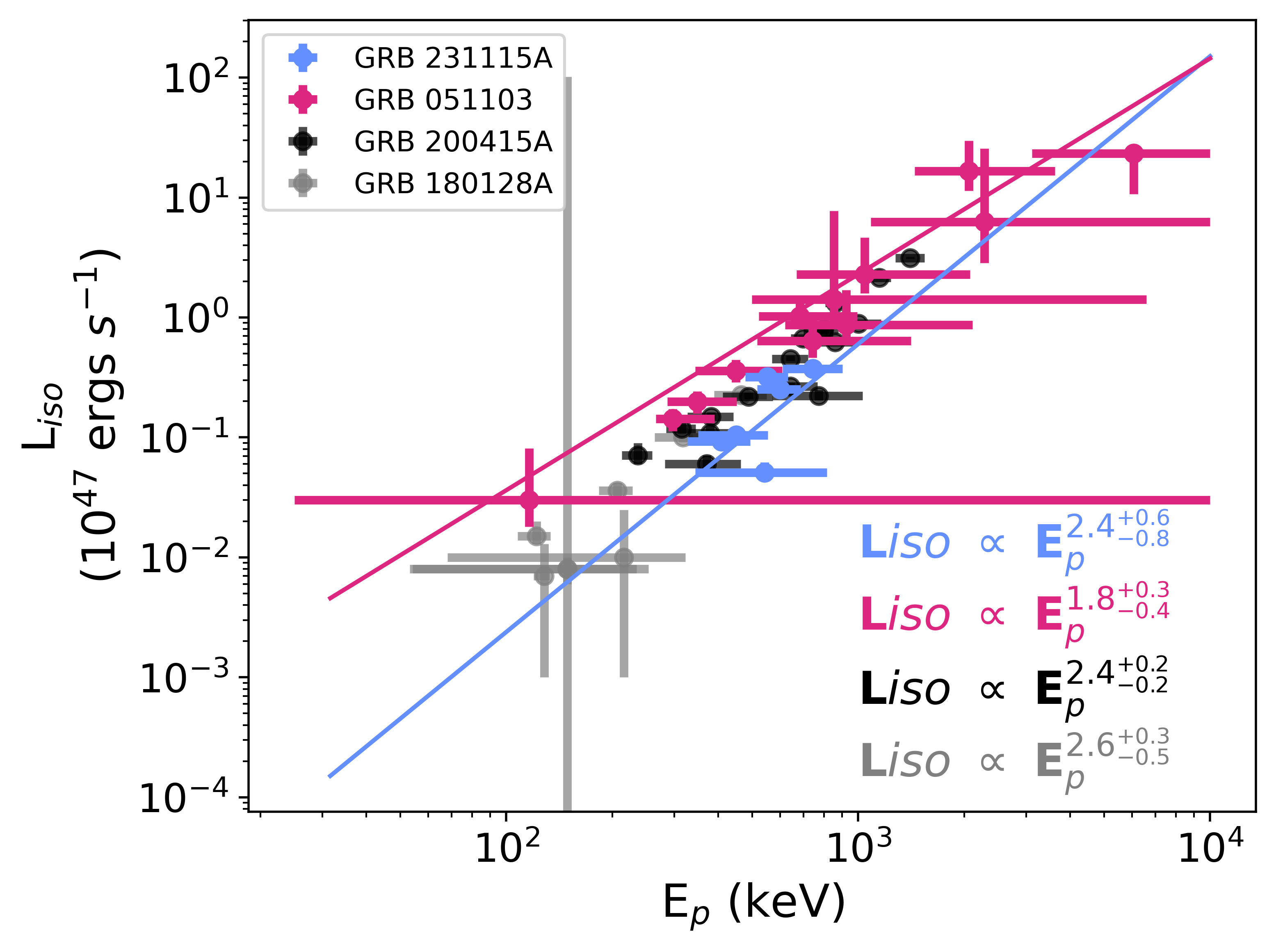}
            
            \textbf{}
            \end{center}
        \end{minipage}
        \begin{minipage}{.5\textwidth}
            \begin{center}
            \includegraphics[width=\linewidth]{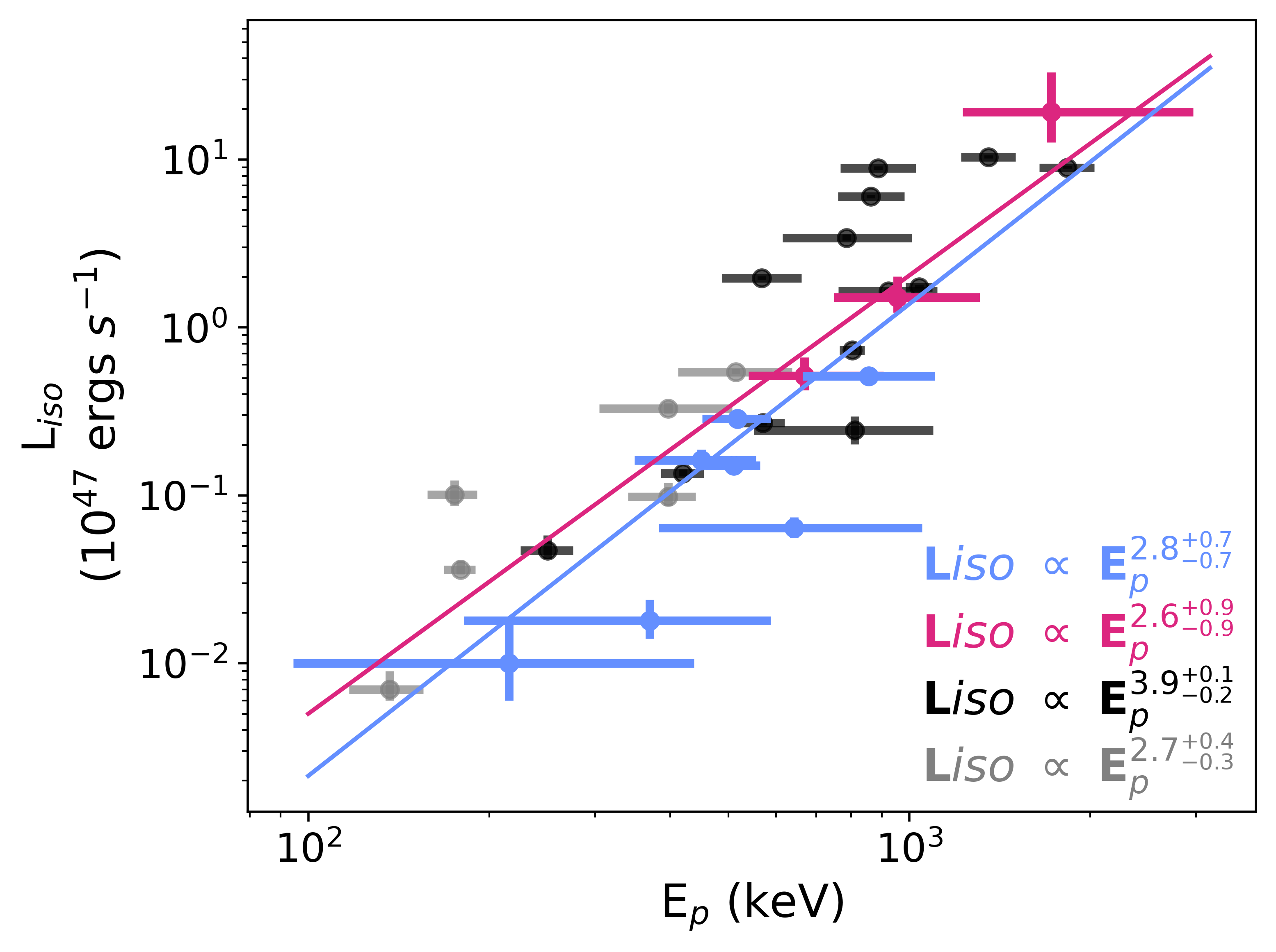}
            
            \textbf{}
            \end{center}
        \end{minipage}\hfill 
        \caption{$L_{\mathrm{iso}}$--$E_{\mathrm{p}}$ relation using different timing intervals. \textit{Left:} Correlation between $L_{\mathrm{iso}}$ and $E_{\mathrm{p}}$ for the four MGF candidates, GRB\,231115A (light blue), GRB\,051103 (pink), GRB\,200415A (black), and GRB\,180128A (gray), revealing an approximate $L_{\mathrm{iso}}$ $\propto$ $E_{\mathrm{p}}^{2}$ relationship that is a strong signature of relativistic winds. The temporal binning is uniformly fixed intervals and uses the values in Table\,\ref{tab:GBM_spec_fixed}. \textit{Right:} $L_{\mathrm{iso}}$ and $E_{\mathrm{p}}$ for all four transients over the BB intervals in Tables\,\ref{tab:GBM_spec_BB} and \ref{tab:GBM_spec_other}, omitting the unconstrained $E_{\rm{p}}$ intervals of GRB\,051103. These data were fit using the fitting method described in Appendix\,\ref{sec:fitting_meth}, which is able to fit data with asymmetric uncertainties. All fit uncertainties and uncertainty bars are at the 90\% confidence level.}
        \label{fig:GBM_Ep_evol_Liso}
    \end{figure*}
    
    Nonuniformity of the expanding wind naturally yields fluctuations in the Doppler elements, and these are reflected in both the light curve, which has a flux (luminosity) tracer $L_{\mathrm{iso}}$, and the spectroscopy, for which we used $E_\mathrm{p}$ as a hardness marker. In Fig.~\ref{fig:GBM_Ep_evol_Liso} we illustrate that these two quantities are fairly tightly correlated for GRB\,231115A and prior MGFs approximately via the $L_\mathrm{iso} \propto E_{\mathrm{p}}^{\beta}$ form listed in Eq.~(\ref{eq:Liso_Ep_correl}). \cite{2021Natur.589..207R} emphasized that theoretically, the index $\beta$ should be tightly coupled to the dual relativistic elements of Doppler boosting (influencing $E_\mathrm{p} \propto \Gamma $) and Doppler beaming (enhancing $L_\mathrm{iso}$). As was detailed in \cite{Trigg2024A&A...687A.173T}, the actual value of the index $\beta$ should depend on the temporal sampling of the light curve, ranging from $\beta \sim 2$ for approximately uniform time bins (see Fig.~\ref{fig:GBM_Ep_evol_Liso}a), to $\beta \sim 4$ for BB time intervals (see Fig.~\ref{fig:GBM_Ep_evol_Liso}b)\footnote{A detailed description of our fitting method can be found in Appendix\,\ref{sec:fitting_meth}}. The physical origin of these two extremes is linked to the size of the look-back surface within the relativistic wind that is sampled during the time intervals. The BB choice yields partial samplings of smaller surface sizes that capture more of the Doppler flux beaming character $L_\mathrm{iso} \propto \Gamma^4$. Performing spectroscopy on uniform time bin selections tends to smear out the surface solid angle sampling fluctuations, emphasizing the photon energy boosting and time dilation characteristics, generating $L_\mathrm{iso} \propto \Gamma^2$ \citep{Trigg2024A&A...687A.173T}. For GRB\,231115A, the reality is between these two extremes, as it is for the other MGFs addressed in Fig.~\ref{fig:GBM_Ep_evol_Liso} and discussed in \cite{Trigg2024A&A...687A.173T}. Yet the fact that a steeper $L_\mathrm{iso} - E_\mathrm{p}$ correlation is evinced for the BB choice does indicate that the physical angular extent of the wind is modest or small, commensurate with expectations of a wind anchored to open field lines emerging from the magnetar's polar surface.

    Based on the saturation limit discussed in Sect.\,\ref{sec:persist}, the prospect of detecting the persistent emission from a magnetar following a MGF, whether during quiescence or an outburst, is slim at distances $\gtrsim3.5$~Mpc. This situation slightly improves with the next-generation X-ray telescopes. For instance, the sensitivity limit for AXIS, a NASA Probe mission selected for Phase A \citep{AXIS-10.1117/12.2677468}, is about one to two orders of magnitude lower than \textit{Chandra's}, depending on the spectral shape of the underlying source population, a 250~ks exposure with AXIS reaches a limit of $\approx 10^{-16}$~erg~s$^{-1}$~cm$^{-2}$ for a $5\sigma$ detection, which translates to $10^{35}$~erg~s$^{-1}$ at 3.5~Mpc \citep{safiharb23:axis}. The prospect of identifying the magnetar, through either spectral or temporal analyses, is better for nearer galaxies such as Andromeda or M33. However, this limited volume constrains the number density of MGFs. It is clear that identifying the persistent counterpart to a candidate extragalactic MGF (in either quiescence or outburst) will require missions like AXIS and flagship X-ray missions such as Lynx or Athena; the launch of Athena is planned for late 2030. Additionally, an intentionally developed IPN could achieve arcsecond localization directly from the event spike.

    Turning to the radio observations, the constraints mentioned in Sect.\,\ref{sec:intro} are highly restrictive, with limits ranging from $<10^{-11}$ to $10^{-9}$ in radio-to-gamma-ray fluence given a flat radio spectral index and band extent of approximately a few times $10^8$ Hz. It remains uncertain whether MGFs can produce prompt FRBs, as the statistics of FRBs resemble those of short bursts rather than MGFs \citep[although shock models favor MGFs; see][]{2010vaoa.conf..129P,Lyubarsky14,Beloborodov17,Metzger+19}. FRB(s) associated with the SGR 1806--20 MGF in 2004 would have been detected by Murriyang \citep{2016ApJ...827...59T}, suggesting that MGFs do not generically result in FRBs. However, from the limited Galactic examples available, MGFs occur during active states of the magnetar, often coinciding with numerous short bursts. Therefore, FRBs could potentially be correlated with MGFs, though not directly caused by them \citep{Ridnaia2024MNRAS.527.5580R}. Patchy observations from 2020-2022 searching for FRBs in M82 yielded limits of $1.2\times 10^{28}$\,erg\,s$^{-1}$\,Hz$^{-1}$ \citep{2024MNRAS.528.6340P}.
    
\section{Conclusion} \label{sec:conclusion}

    The detection and analysis of GRB\,231115A provide compelling evidence that it is an extragalactic MGF originating from the starburst galaxy M82. This burst gives a fully self-consistent picture based on the model discussed in Sect.\,\ref{sec:discuss}. The observed spectral evolution closely mirrors the characteristics of known MGFs from other galaxies, such as NGC\,253, supporting the classification of GRB\,231115A as a MGF. The comprehensive gamma-ray analysis conducted with \fgbm data provides us with significant insights into the burst's properties, including its high peak energy and complex temporal structure. The association of GRB\,231115A with M82 is further strengthened by the high Bayes factor, suggesting a very low probability of a chance alignment with a cosmological neutron star merger.
    
    The detection of high-energy photons at twice the threshold for $\gamma\gamma$ pair creation confirms a lower limit on the bulk Lorentz factor ($\Gamma \gtrsim 2$) of the outflowing plasma, consistent with expectations for relativistic winds from magnetars, and is comparable to observations from previous MGFs such as GRB\,200415A and GRB\,180128A. Moreover, the MVT indicates rapid flux changes typical of MGFs. This short timescale aligns with the smaller Lorentz factors expected for MGFs compared to cosmological GRBs, further supporting our classification.
    
    This discovery not only adds to the growing list of extragalactic MGF candidates but also emphasizes the importance of prompt and coordinated multiwavelength follow-up observations for understanding these rare and energetic events. The ability to detect and accurately localize such MGFs in distant galaxies opens up new avenues for studying magnetars and their environments outside our Galaxy. The detectability of MGF tails at extragalactic distances is crucial, as they provide unambiguous evidence of magnetar central engines. Simulations show that the detectability of the tail depends significantly on the re-pointing time and instrument background rates. The limitations in detecting and characterizing MGFs with current instruments underscore the importance of new technologies and coordinated multiwavelength observations. Proposed missions and next-generation X-ray satellites are critical for advancing our understanding of these phenomena.
    
    The detection of GRB\,231115A and its unambiguous localization to M82 underscores the contribution of MGFs to the population of short GRBs. Future advancements in observational technology and methodologies, such as rapid automated re-pointing and increased X-ray sensitivity, will likely enhance our capacity to identify and study these phenomena, thereby furthering our understanding of the mechanisms driving short GRBs and the role of magnetars in the cosmos.

\begin{acknowledgements}

We thank the reviewer for their insightful comments and suggestions that significantly improved the quality of this manuscript. AT, EB, MN, and OJR acknowledge NASA support under award 80NSSC21K2038. MGB thanks NASA for support under grants 80NSSC22K0777 and 80NSSC22K1576. Z.W. acknowledges support by NASA under award number 80GSFC21M0002. The USRA coauthors gratefully acknowledge NASA funding through cooperative agreement 80NSSC24M0035. BO is supported by the McWilliams Postdoctoral Fellowship at Carnegie Mellon University. This paper contains data obtained at the Wendelstein Observatory of the Ludwig-Maximilians University Munich. The authors greatly acknowledge the assistance of the observer Michael Schmidt (USM) in obtaining the observations. This work was in part funded by the Deutsche Forschungsgemeinschaft (DFG, German Research Foundation) under Germany's Excellence Strategy – EXC-2094 – 390783311. The work of DF and DS was supported by the basic funding program of the Ioffe Institute FFUG-2024-0002.

\end{acknowledgements}

\bibliographystyle{aa}
\bibliography{GRB231115A}

\begin{appendix}

\onecolumn
\section{$L_\mathrm{iso}$--$E_\mathrm{p}$ relation fitting method} \label{sec:fitting_meth}

    In previous analyses of the luminosity-hardness correlation relationship between $L_\mathrm{iso}$ and $E_\mathrm{p}$, such as those in \cite{2021Natur.589..207R} and \cite{Trigg2024A&A...687A.173T}, data fitting used the assumptions of symmetric errors for the dependent ($L_\mathrm{iso}$) and independent variables ($E_\mathrm{p}$), sometimes even ignoring the errors in the $E_\mathrm{p}$ entirely, as is common in least-squares regression. To better constrain the fit values for the $L_\mathrm{iso}$--$E_\mathrm{p}$ relation of GRB\,231115A and GRB\,051103, as well as reevaluate GRB\,200415A, and GRB\,180128A, we used statistical methods that account for asymmetric errors in both variables. This method allows the unbiased determination of probability distributions for the model parameters of interest.

    \begin{figure*}[ht!]
        \begin{minipage}{.39\textwidth}
            \begin{center}
            \includegraphics[width=6cm]{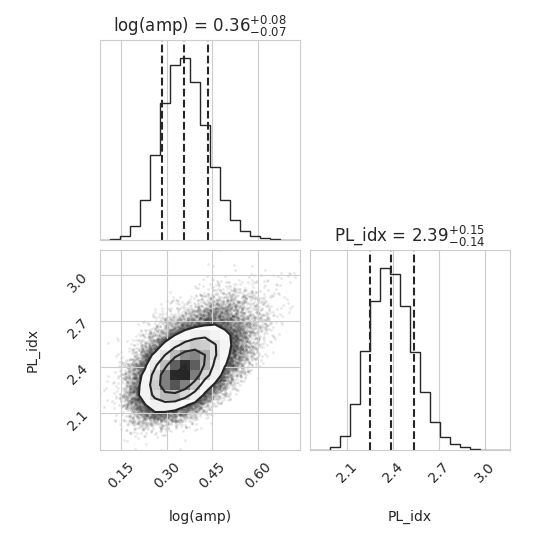}
            
            \textbf{}
            \end{center}
        \end{minipage}
        \begin{minipage}{.6\textwidth}
            \begin{center}
            \includegraphics[width=10cm]{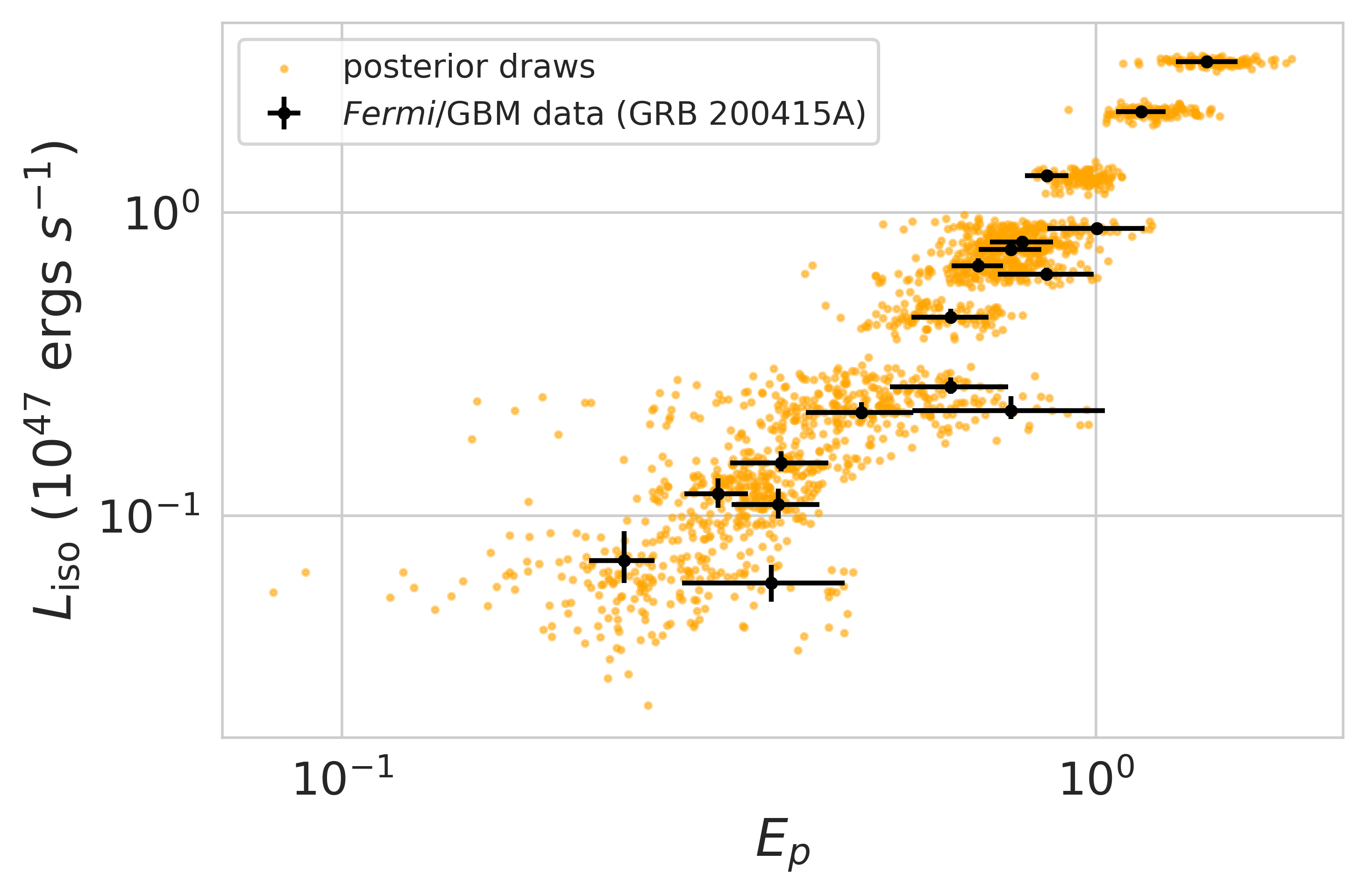}
            
            \textbf{}
            \end{center}
        \end{minipage}\hfill 
        \caption{Example of fitting results\textit{Left:} Corner plot of posterior distributions for the two power-law parameters that describe the relationship between the peak energy and isotropic luminosity. \textit{Right:} Data-resolved spectroscopy of the \fgbm data for GRB\,200415A (black) and the posterior predictive distribution (orange). To compute the posterior predictive distribution, we simulated the data generation process by first picking a set of parameters ($(\{\hat{E}_{p,i,k}\}_{i=1}^{N}, \alpha_k, \beta_k)$) from the posterior, then generating random simulated values for $L_\mathrm{iso, sim}^*$ and $E_{p, \mathrm{sim}^*}$ using the TPN distribution. This figure shows 100 such simulated datasets and enables a convenient comparison of the performance of the model with the data.}
        \label{fig:ep_liso}
    \end{figure*}
    
    Errors on both $E_p$ and $L_\mathrm{iso}$ are heteroscedastic (different for each data point) and asymmetric. To model the asymmetric errors in each dimension, we used a two-piece normal (TPN) distribution as proposed by \citet{fernandez1998}. This distribution corresponds to two half-normal distributions with standard deviations $\sigma_1$ (left-hand side) and $\sigma_2$ (right-hand side), joined at a common mode $\mu$ and renormalized such that the distribution is continuous. This distribution is usually parametrized by a common standard deviation $\sigma$ and a skewness $\lambda$, with the probability density then being 
    
    \begin{equation}\label{eqn:snd}
    p(x, \mu, \sigma, \lambda) = 
    \begin{cases}
    \frac{2\lambda}{(1 + \lambda^2)\sigma} \phi\left(\frac{\lambda(x - \mu)}{\sigma}\right) \, , \; \mathrm{if} \, x < \mu; \\
    \frac{2\lambda}{(1 + \lambda^2)\sigma} \phi\left(\frac{x - \mu}{\lambda\sigma}\right) \, , \; \mathrm{if} \, x \geq \mu \, ,
    \end{cases}
    \end{equation}
    
    \noindent where 
    
    \begin{eqnarray}\label{eqn:tpn_conversion}
    \sigma &=& \sigma_1 \\ \nonumber
    \lambda &=& \sqrt{\sigma_2 / \sigma_1} \nonumber \;.
    \end{eqnarray}
    
    \noindent These two equations enable the straightforward conversion of asymmetric errors as reported in the astronomical literature into a single standard deviation and skewness for use with the TPN. 
    
    To account for the presence of errors in both variables, we built an errors-in-variables model. Our data consists of pairs of observed values $\{E_p^*, L_\mathrm{iso}^*\}$. We assumed that the observed $E_p^*$ are random variables drawn from a TPN distribution with a mean given by a set of unknown true $\hat{E}_p$ values.
    To identify the luminosity-hardness correlation, we assumed a standard power-law relationship between the variables of the form
    \begin{equation}
    \hat{L}_\mathrm{iso} = \alpha \hat{E}_p^{\beta} \quad .
     \label{eq:Liso_Ep_correl}
    \end{equation}
    We note that this relationship holds between the (unknown) true peak energy and isotropic luminosity, not the observed quantities. Whereas a standard likelihood assumes that the $\hat{E}_p$ are known, the errors-in-variables model infers them along with the parameters of the relationship, such that the total number of parameters to be inferred becomes $(\{\hat{E}_{p,i}\}_{i=1}^{N}, \alpha, \beta )$, for $N$ data points, a power-law amplitude $\alpha$ and a power-law index $\beta$. We infer the parameters in a Bayesian framework:
    \begin{equation}
    p(\alpha, \beta, \{\hat{E}_{p,i}\}_{i=1}^{N} | E_p^*, L_\mathrm{iso}^*) \propto p(L_\mathrm{iso}^* | \hat{E}_p, \alpha, \beta) p(E_p^* | \hat{E}_p) p(\alpha, \beta, \hat{E}_p) \; .
     \label{eq:Bayesian_param}
    \end{equation}
    \noindent Intuitively, this model parametrizes the data generation process: a power-law relationship exists between the unknown true peak energies and the unknown true isotropic luminosities. We first parametrized the relationship between the true peak energies and observed peak energies using the TPN distribution, and drew from it as well as the priors for the other parameters $\alpha$ and $\beta$. We can assume conditional independence and as such the priors for all parameters can be written independently:

    \[
    p(\alpha, \beta, \hat{E}_p) = p(\alpha) p(\beta) p(\hat{E}_p) \; .
    \]

    The observed isotropic luminosities are then governed by a power-law relationship between the sampled true $\hat{E}_p$ and true, unknown $\hat{L}_\mathrm{iso}$, given parameters $\alpha$ and $\beta$, and compared to the observed $L^*_\mathrm{iso}$ through a likelihood $p(L_\mathrm{iso}^* | \hat{E}_p, \alpha, \beta) \sim \mathrm{TPN}(\hat{L}_\mathrm{iso}, \sigma_L, \lambda_L)$, where $\sigma_L$ and $\lambda_L$ are the standard deviation and skewness for $L^*_\mathrm{iso}$ as defined using Eq. \ref{eqn:tpn_conversion}.
    
    This model can be sampled with Markov chain Monte Carlo in order to infer the true $\hat{E}_p$ and the power-law parameters $\alpha$ and $\beta$. To do so, one first samples $\hat{E}_p$ from a prior distribution, and then computes a likelihood $p(E_p^* | \hat{E}_p)$ for the observed peak energy given the true peak energy using the TPN distribution. In a second step, the $\hat{E}_p$ values are used to compute the proposed power-law relationship, together with $\alpha$ and $\beta$, also sampled from a prior. This relationship is then compared to the observed $L_\mathrm{iso}^*$, again using a TPN distribution to obtain the posterior probability for the parameters. This model enables us to effectively take into account the asymmetric uncertainties on both peak energy and isotropic luminosity.
    
    We set uninformative, flat priors on all parameters, where the prior for peak energies are defined as $\hat{E}_p \sim \mathcal{U}(0 \,\mathrm{keV}, 15\,\mathrm{keV})$, the prior on the power-law index is $\beta \sim \mathcal{U}(1,4)$ and the prior on the natural logarithm of the amplitude is $\log(\alpha) \sim \mathcal{U}(-2, 4)$, where $\mathcal{U}(a,b)$ denotes a uniform distribution between lower boundary $a$ and upper boundary $b$.

    We implemented the model in the probabilistic programming language \texttt{numpyro} \citep{phan2019composable,bingham2019pyro} and used Hamiltonian Monte Carlo with the No U-Turn Sampler \citep[NUTS;][]{10.5555/2627435.2638586} to sample the posteriors for all parameters. We ran six Hamiltonian Monte Carlo chains with 4,000 warmup steps and 10,000 sampling steps. We assessed convergence using autocorrelation time and the Gelman-Rubin statistic and find that all chains are well converged ($T_\mathrm{GR} \leq 1.001$ for all parameters). 
    
    Figure \ref{fig:ep_liso} shows the results of the modeling for GRB\,200415A. We chose this burst as it has good count statistics and the luminosity-hardness correlation has been derived by several previous studies \citep{2021Natur.589..207R, Trigg2024A&A...687A.173T}. The posterior for the parameters is well constrained and single-peaked, with a power-law index of $\beta = 2.39^{+0.15}_{-0.14}$, marginalized over the uncertainties in $E_p$. Similarly, the posterior predictive distribution mirrors the observed data well, suggesting that the model overall can represent the data effectively.

\onecolumn
\begin{table*}[ht!]
\section{Time-resolved spectral analyses using fixed-time intervals} \label{time-res_fixed}
    \caption{{Time-resolved analyses: Fixed intervals.}}
    \label{tab:GBM_spec_fixed}
    \centering
    \begin{tabular}{cccccc}
    \hline\hline
    Time & $E_{\mathrm{p}}$ & $\alpha$ & Energy Flux ($\cal{F}$) & $L_{\mathrm{iso}}$ &  $E_{\mathrm{iso}}$\\
    (ms) & (keV) &  & ($\times10^{-6}$~ergs~s$^{-1}$~cm$^{-2}$) &($\times10^{45}$~erg$\cdot$~s$^{-1}$) &  ($\times10^{44}$~erg)\\
    \hline
    \multicolumn{6}{c}{GRB\,231115A}\\
    -16:-8 & 550 {\raisebox{0.5ex}{\tiny$\substack{+80 \\ -70}$}} & 0.5  {\raisebox{0.5ex}{\tiny$\substack{+0.5 \\ -0.5}$}} & 22  {\raisebox{0.5ex}{\tiny$\substack{+2 \\ -2}$}} & 32  {\raisebox{0.5ex}{\tiny$\substack{+3 \\ -2}$}} & 2.7  {\raisebox{0.5ex}{\tiny$\substack{+0.2 \\ -0.12}$}} \\ 
    -8:0 & 740 {\raisebox{0.5ex}{\tiny$\substack{+160 \\ -130}$}} & 0.1  {\raisebox{0.5ex}{\tiny$\substack{+0.4 \\ -0.4}$}} & 25  {\raisebox{0.5ex}{\tiny$\substack{+2 \\ -2}$}} & 37  {\raisebox{0.5ex}{\tiny$\substack{+2 \\ -1}$}} & 3.14  {\raisebox{0.5ex}{\tiny$\substack{+0.14 \\ -0.11}$}} \\ 
    0:8 & 600 {\raisebox{0.5ex}{\tiny$\substack{+90 \\ -80}$}} & 0.9  {\raisebox{0.5ex}{\tiny$\substack{+0.5 \\ -0.6}$}} & 17  {\raisebox{0.5ex}{\tiny$\substack{+2 \\ -2.}$}} & 25.2  {\raisebox{0.5ex}{\tiny$\substack{+2.3 \\ -1.1}$}} & 2.19  {\raisebox{0.5ex}{\tiny$\substack{+0.20 \\ -0.012}$}} \\ 
    8:16 & 410 {\raisebox{0.5ex}{\tiny$\substack{+90 \\ -80}$}} & 0.8  {\raisebox{0.5ex}{\tiny$\substack{+1.5 \\ -1.1}$}} & 6  {\raisebox{0.5ex}{\tiny$\substack{+1 \\ -2}$}} & 9.3  {\raisebox{0.5ex}{\tiny$\substack{+1.4 \\ -0.7}$}} & 0.85  {\raisebox{0.5ex}{\tiny$\substack{+0.11 \\ -0.07}$}} \\ 
    16:24 & 500 {\raisebox{0.5ex}{\tiny$\substack{+100 \\ -100}$}} & 0.7  {\raisebox{0.5ex}{\tiny$\substack{+1.2 \\ -1.0}$}} & 7  {\raisebox{0.5ex}{\tiny$\substack{+2 \\ -2}$}} & 10  {\raisebox{0.5ex}{\tiny$\substack{+2 \\ -1}$}} & 0.94  {\raisebox{0.5ex}{\tiny$\substack{+0.12 \\ -0.08}$}} \\ 
    24:44 & 500 {\raisebox{0.5ex}{\tiny$\substack{+300 \\ -200}$}} & -0.2  {\raisebox{0.5ex}{\tiny$\substack{+0.9 \\ -0.7}$}} & 3.5  {\raisebox{0.5ex}{\tiny$\substack{+1.0 \\ -0.8}$}} & 5.1  {\raisebox{0.5ex}{\tiny$\substack{+1.1 \\ -0.7}$}} & 1.21  {\raisebox{0.5ex}{\tiny$\substack{+0.20 \\ -0.14}$}} \\ 
    \hline\hline
    \multicolumn{6}{c}{GRB\,200415A}\\
    -8:0 & 1190 {\raisebox{0.5ex}{\tiny$\substack{+120 \\ -110}$}} & -0.3  {\raisebox{0.5ex}{\tiny$\substack{+0.1 \\ -0.1}$}} & 276  {\raisebox{0.5ex}{\tiny$\substack{+8 \\ -8}$}} & 404  {\raisebox{0.5ex}{\tiny$\substack{+9 \\ -6}$}} & 33.0  {\raisebox{0.5ex}{\tiny$\substack{+0.8 \\ -0.5}$}} \\ 
    0:8 & 1400 {\raisebox{0.5ex}{\tiny$\substack{+140 \\ -130}$}} & 0.1  {\raisebox{0.5ex}{\tiny$\substack{+0.1 \\ -0.1}$}} & 214  {\raisebox{0.5ex}{\tiny$\substack{+7 \\ -7}$}} & 314  {\raisebox{0.5ex}{\tiny$\substack{+8 \\ -4}$}} & 25.6  {\raisebox{0.5ex}{\tiny$\substack{+0.7 \\ -0.4}$}} \\ 
    8:16 & 1150 {\raisebox{0.5ex}{\tiny$\substack{+90 \\ -90}$}} & 0.6  {\raisebox{0.5ex}{\tiny$\substack{+0.1 \\ -0.1}$}} & 146  {\raisebox{0.5ex}{\tiny$\substack{+7 \\ -8}$}} & 215  {\raisebox{0.5ex}{\tiny$\substack{+7 \\ -5}$}} & 17.8  {\raisebox{0.5ex}{\tiny$\substack{+0.6 \\ -0.4}$}} \\ 
    16:24 & 860 {\raisebox{0.5ex}{\tiny$\substack{+60 \\ -60}$}} & 1.1  {\raisebox{0.5ex}{\tiny$\substack{+0.2 \\ -0.2}$}} & 90  {\raisebox{0.5ex}{\tiny$\substack{+5 \\ -6}$}} & 132  {\raisebox{0.5ex}{\tiny$\substack{+5 \\ -4}$}} & 11.0  {\raisebox{0.5ex}{\tiny$\substack{+0.5 \\ -0.3}$}} \\ 
    24:32 & 1000 {\raisebox{0.5ex}{\tiny$\substack{+200 \\ -140}$}} & 0.03  {\raisebox{0.5ex}{\tiny$\substack{+0.20 \\ -0.20}$}} & 61  {\raisebox{0.5ex}{\tiny$\substack{+3 \\ -3}$}} & 89  {\raisebox{0.5ex}{\tiny$\substack{+3 \\ -3}$}} & 7.3  {\raisebox{0.5ex}{\tiny$\substack{+0.2 \\ -0.2}$}} \\ 
    32:40 & 770 {\raisebox{0.5ex}{\tiny$\substack{+80 \\ -70}$}} & 0.7  {\raisebox{0.5ex}{\tiny$\substack{+0.2 \\ -0.2}$}} & 52  {\raisebox{0.5ex}{\tiny$\substack{+4 \\ -4}$}} & 76  {\raisebox{0.5ex}{\tiny$\substack{+4 \\ -3}$}} & 6.3  {\raisebox{0.5ex}{\tiny$\substack{+0.4 \\ -0.2}$}} \\ 
    40:48 & 800 {\raisebox{0.5ex}{\tiny$\substack{+80 \\ -80}$}} & 0.7  {\raisebox{0.5ex}{\tiny$\substack{+0.2 \\ -0.2}$}} & 54  {\raisebox{0.5ex}{\tiny$\substack{+4 \\ -4}$}} & 80  {\raisebox{0.5ex}{\tiny$\substack{+5 \\ -3}$}} & 6.7  {\raisebox{0.5ex}{\tiny$\substack{+0.3 \\ -0.3}$}} \\ 
    48:56 & 700 {\raisebox{0.5ex}{\tiny$\substack{+60 \\ -50}$}} & 1.2  {\raisebox{0.5ex}{\tiny$\substack{+0.2 \\ -0.3}$}} & 46  {\raisebox{0.5ex}{\tiny$\substack{+3 \\ -4}$}} & 67  {\raisebox{0.5ex}{\tiny$\substack{+4 \\ -2}$}} & 6.0  {\raisebox{0.5ex}{\tiny$\substack{+0.3 \\ -0.2}$}} \\ 
    56:64 & 860 {\raisebox{0.5ex}{\tiny$\substack{+130 \\ -120}$}} & 0.2  {\raisebox{0.5ex}{\tiny$\substack{+0.2 \\ -0.3}$}} & 43  {\raisebox{0.5ex}{\tiny$\substack{+3 \\ -3}$}} & 63  {\raisebox{0.5ex}{\tiny$\substack{+4 \\ -2}$}} & 5.2  {\raisebox{0.5ex}{\tiny$\substack{+0.3 \\ -0.2}$}} \\ 
    64:72 & 640 {\raisebox{0.5ex}{\tiny$\substack{+80 \\ -70}$}} & 0.6  {\raisebox{0.5ex}{\tiny$\substack{+0.3 \\ -0.4}$}} & 31  {\raisebox{0.5ex}{\tiny$\substack{+3 \\ -3}$}} & 45  {\raisebox{0.5ex}{\tiny$\substack{+3 \\ -2}$}} & 3.8  {\raisebox{0.5ex}{\tiny$\substack{+0.2 \\ -0.1}$}} \\ 
    72:80 & 640 {\raisebox{0.5ex}{\tiny$\substack{+120 \\ -110}$}} & 0.2  {\raisebox{0.5ex}{\tiny$\substack{+0.4 \\ -0.5}$}} & 18  {\raisebox{0.5ex}{\tiny$\substack{+2 \\ -2}$}} & 26.6  {\raisebox{0.5ex}{\tiny$\substack{+2.4 \\ -1.4}$}} & 2.3  {\raisebox{0.5ex}{\tiny$\substack{+0.2 \\ -0.1}$}} \\ 
    80:88 & 490 {\raisebox{0.5ex}{\tiny$\substack{+80 \\ -80}$}} & 0.3  {\raisebox{0.5ex}{\tiny$\substack{+0.6 \\ -0.6}$}} & 15  {\raisebox{0.5ex}{\tiny$\substack{+2 \\ -2}$}} & 21.9  {\raisebox{0.5ex}{\tiny$\substack{+1.6 \\ -1.0}$}} & 2.0  {\raisebox{0.5ex}{\tiny$\substack{+0.1 \\ -0.1}$}} \\ 
    88:96 & 380 {\raisebox{0.5ex}{\tiny$\substack{+60 \\ -60}$}} & 0.6  {\raisebox{0.5ex}{\tiny$\substack{+1.3 \\ -0.8}$}} & 10.2  {\raisebox{0.5ex}{\tiny$\substack{+1.3 \\ -1.5}$}} & 14.9  {\raisebox{0.5ex}{\tiny$\substack{+1.5 \\ -0.9}$}} & 1.28  {\raisebox{0.5ex}{\tiny$\substack{+0.11 \\ -0.06}$}} \\ 
    96:104 & 800 {\raisebox{0.5ex}{\tiny$\substack{+300 \\ -200}$}} & -0.12  {\raisebox{0.5ex}{\tiny$\substack{+0.43 \\ -0.47}$}} & 15  {\raisebox{0.5ex}{\tiny$\substack{+2 \\ -2}$}} & 22  {\raisebox{0.5ex}{\tiny$\substack{+3 \\ -2}$}} & 2.0  {\raisebox{0.5ex}{\tiny$\substack{+0.2 \\ -0.2}$}} \\ 
    104:112 & 380 {\raisebox{0.5ex}{\tiny$\substack{+50 \\ -50}$}} & 1.4  {\raisebox{0.5ex}{\tiny$\substack{+1.5 \\ -1.2}$}} & 7.4  {\raisebox{0.5ex}{\tiny$\substack{+1.5 \\ -9.1}$}} & 10.9  {\raisebox{0.5ex}{\tiny$\substack{+1.8 \\ -1.1}$}} & 1.01  {\raisebox{0.5ex}{\tiny$\substack{+0.11 \\ -0.10}$}} \\ 
    112:120 & 320 {\raisebox{0.5ex}{\tiny$\substack{+30 \\ -30}$}} & 1.9  {\raisebox{0.5ex}{\tiny$\substack{+3.2 \\ -1.3}$}} & 8  {\raisebox{0.5ex}{\tiny$\substack{+2 \\ -2}$}} & 11.8  {\raisebox{0.5ex}{\tiny$\substack{+2.0 \\ -1.1}$}} & 1.08  {\raisebox{0.5ex}{\tiny$\substack{+0.14 \\ -0.10}$}} \\ 
    120:128 & 240 {\raisebox{0.5ex}{\tiny$\substack{+20 \\ -20}$}} & 2.2  {\raisebox{0.5ex}{\tiny$\substack{+1.2 \\ -0.8}$}} & 4.9  {\raisebox{0.5ex}{\tiny$\substack{+2.0 \\ -1.1}$}} & 7.1  {\raisebox{0.5ex}{\tiny$\substack{+1.5 \\ -1.2}$}} & 0.69  {\raisebox{0.5ex}{\tiny$\substack{+0.14 \\ -0.08}$}} \\ 
    128:136 & 370 {\raisebox{0.5ex}{\tiny$\substack{+90 \\ -90}$}} & 0.7  {\raisebox{0.5ex}{\tiny$\substack{+2.3 \\ -1.2}$}} & 4.1  {\raisebox{0.5ex}{\tiny$\substack{+1.0 \\ -1.3}$}} & 6.0  {\raisebox{0.5ex}{\tiny$\substack{+0.9 \\ -0.7}$}} & 0.57  {\raisebox{0.5ex}{\tiny$\substack{+0.08 \\ -0.06}$}} \\ 
    \hline\hline
    \multicolumn{6}{c}{GRB\,180128A}\\
    -16:-6 & 460 {\raisebox{0.5ex}{\tiny$\substack{+80 \\ -80}$}} & 0.2  {\raisebox{0.5ex}{\tiny$\substack{+0.6 \\ -0.5}$}} & 15.4  {\raisebox{0.5ex}{\tiny$\substack{+1.4 \\ -1.5}$}} & 22.5  {\raisebox{0.5ex}{\tiny$\substack{+1.5 \\ -0.9}$}} & 2.39  {\raisebox{0.5ex}{\tiny$\substack{+0.15 \\ -0.10}$}} \\ 
    -6:4 & 320 {\raisebox{0.5ex}{\tiny$\substack{+60 \\ -50}$}} & 0.6  {\raisebox{0.5ex}{\tiny$\substack{+1.3 \\ -0.7}$}} & 6.8  {\raisebox{0.5ex}{\tiny$\substack{+1.2 \\ -1.2}$}} & 10.0  {\raisebox{0.5ex}{\tiny$\substack{+1.2 \\ -0.9}$}} & 1.11  {\raisebox{0.5ex}{\tiny$\substack{+0.14 \\ -0.07}$}} \\ 
    4:14 & 210 {\raisebox{0.5ex}{\tiny$\substack{+20 \\ -20}$}} & 3.3  {\raisebox{0.5ex}{\tiny$\substack{+1.4 \\ -1.0}$}} & 2.5  {\raisebox{0.5ex}{\tiny$\substack{+0.7 \\ -0.6}$}} & 3.6  {\raisebox{0.5ex}{\tiny$\substack{+0.7 \\ -0.5}$}} & 0.42  {\raisebox{0.5ex}{\tiny$\substack{+0.06 \\ -0.04}$}} \\ 
    14:24 & 120 {\raisebox{0.5ex}{\tiny$\substack{+12 \\ -14}$}} & 4.3  {\raisebox{0.5ex}{\tiny$\substack{+0.8 \\ -0.6}$}} & 1.0  {\raisebox{0.5ex}{\tiny$\substack{+0.4 \\ -0.3}$}} & 1.5  {\raisebox{0.5ex}{\tiny$\substack{+0.5 \\ -0.3}$}} & 0.19  {\raisebox{0.5ex}{\tiny$\substack{+0.04 \\ -0.03}$}} \\ 
    24:34 & 150 {\raisebox{0.5ex}{\tiny$\substack{+90 \\ -90}$}} & 0.4  {\raisebox{0.5ex}{\tiny$\substack{+3.0 \\ -0.8}$}} & 0.6  {\raisebox{0.5ex}{\tiny$\substack{+11360.0 \\ -29.5}$}} & 0.8  {\raisebox{0.5ex}{\tiny$\substack{+100.0 \\ -94.3}$}} & 1200  {\raisebox{0.5ex}{\tiny$\substack{+1200 \\ -850}$}} \\ 
    34:44 & 200 {\raisebox{0.5ex}{\tiny$\substack{+100 \\ -100}$}} & 0.1  {\raisebox{0.5ex}{\tiny$\substack{+2.7 \\ -0.8}$}} & 0.6  {\raisebox{0.5ex}{\tiny$\substack{+0.2 \\ -0.3}$}} & 0.8  {\raisebox{0.5ex}{\tiny$\substack{+0.3 \\ -0.2}$}} & 0.11  {\raisebox{0.5ex}{\tiny$\substack{+0.03 \\ -0.02}$}} \\ 
    44:54 & 1280 {\raisebox{0.5ex}{\tiny$\substack{+5 \\ -9}$}} & 20  {\raisebox{0.5ex}{\tiny$\substack{+6 \\ -1}$}} & 0.5  {\raisebox{0.5ex}{\tiny$\substack{+0.8 \\ -0.3}$}} & 0.7  {\raisebox{0.5ex}{\tiny$\substack{+0.9 \\ -0.6}$}} & 0.14  {\raisebox{0.5ex}{\tiny$\substack{+0.07 \\ -0.05}$}} \\ 
    54:64 & 220 {\raisebox{0.5ex}{\tiny$\substack{+110 \\ -150}$}} & 0.9  {\raisebox{0.5ex}{\tiny$\substack{+10.3 \\ -1.4}$}} & 0.7  {\raisebox{0.5ex}{\tiny$\substack{+1.5 \\ -0.8}$}} & 1.0  {\raisebox{0.5ex}{\tiny$\substack{+1.7 \\ -0.9}$}} & 0.24  {\raisebox{0.5ex}{\tiny$\substack{+0.15 \\ -0.10}$}} \\  
    \hline\hline
    \multicolumn{6}{c}{GRB\,051103}\\
    -16:-4 & 120 {\raisebox{0.5ex}{\tiny$\substack{+9900 \\ -100}$}} & -1.1  {\raisebox{0.5ex}{\tiny$\substack{+6.1 \\ -0.9}$}} & 2.1  {\raisebox{0.5ex}{\tiny$\substack{+3.3 \\ 0.9}$}} & 3.0  {\raisebox{0.5ex}{\tiny$\substack{+4.8 \\ -1.4}$}} & 0.4  {\raisebox{0.5ex}{\tiny$\substack{+0.6 \\ -0.2}$}} \\ 
    -4:8 & 2100 {\raisebox{0.5ex}{\tiny$\substack{+1600 \\ -600}$}} & -0.2  {\raisebox{0.5ex}{\tiny$\substack{+0.2 \\ -0.2}$}} & 110  {\raisebox{0.5ex}{\tiny$\substack{+900 \\ 400}$}} & 1670  {\raisebox{0.5ex}{\tiny$\substack{+1400 \\ -500}$}} & 200  {\raisebox{0.5ex}{\tiny$\substack{+150 \\ -600}$}} \\ 
    8:20 & 6000 {\raisebox{0.5ex}{\tiny$\substack{+4000 \\ -3000}$}} & -0.5  {\raisebox{0.5ex}{\tiny$\substack{+0.2 \\ -0.1}$}} & 1600  {\raisebox{0.5ex}{\tiny$\substack{+110 \\ 880}$}} & 2300  {\raisebox{0.5ex}{\tiny$\substack{+200 \\ -1300}$}} & 280  {\raisebox{0.5ex}{\tiny$\substack{+20 \\ -160}$}} \\ 
    20:32 & 2300 {\raisebox{0.5ex}{\tiny$\substack{+7700 \\ -1200}$}} & 0.04  {\raisebox{0.5ex}{\tiny$\substack{+0.62 \\ -0.38}$}} & 400  {\raisebox{0.5ex}{\tiny$\substack{+1300 \\ 200}$}} & 600  {\raisebox{0.5ex}{\tiny$\substack{+1800 \\ -400}$}} & 90  {\raisebox{0.5ex}{\tiny$\substack{+230 \\ -500}$}} \\ 
    32:44 & 900 {\raisebox{0.5ex}{\tiny$\substack{+5700 \\ -400}$}} & 1.2  {\raisebox{0.5ex}{\tiny$\substack{+3.7 \\ -1.1}$}} & 100  {\raisebox{0.5ex}{\tiny$\substack{+400 \\ 40}$}} & 100  {\raisebox{0.5ex}{\tiny$\substack{+600 \\ -600}$}} & 17  {\raisebox{0.5ex}{\tiny$\substack{+70 \\ -7}$}} \\ 
    44:56 & 1000 {\raisebox{0.5ex}{\tiny$\substack{+1000 \\ -400}$}} & 0.5  {\raisebox{0.5ex}{\tiny$\substack{+1.1 \\ -0.5}$}} & 160  {\raisebox{0.5ex}{\tiny$\substack{+160 \\ 50}$}} & 230  {\raisebox{0.5ex}{\tiny$\substack{+240 \\ -70}$}} & 30  {\raisebox{0.5ex}{\tiny$\substack{+30 \\ -11}$}} \\ 
    56:68 & 700 {\raisebox{0.5ex}{\tiny$\substack{+300 \\ -200}$}} & 0.5  {\raisebox{0.5ex}{\tiny$\substack{+0.8 \\ -0.5}$}} & 70  {\raisebox{0.5ex}{\tiny$\substack{+25 \\ 14}$}} & 100  {\raisebox{0.5ex}{\tiny$\substack{+30 \\ -20}$}} & 12  {\raisebox{0.5ex}{\tiny$\substack{+5 \\ -2}$}} \\ 
    68:80 & 900 {\raisebox{0.5ex}{\tiny$\substack{+1200 \\ -300}$}} & -0.3  {\raisebox{0.5ex}{\tiny$\substack{+0.4 \\ -0.3}$}} & 60  {\raisebox{0.5ex}{\tiny$\substack{+60 \\ 20}$}} & 90  {\raisebox{0.5ex}{\tiny$\substack{+80 \\ -30}$}} & 10  {\raisebox{0.5ex}{\tiny$\substack{+11 \\ -3}$}} \\ 
    80:92 & 700 {\raisebox{0.5ex}{\tiny$\substack{+700 \\ -200}$}} & 0.2  {\raisebox{0.5ex}{\tiny$\substack{+1.0 \\ -0.5}$}} & 43  {\raisebox{0.5ex}{\tiny$\substack{+31 \\ 11}$}} & 64  {\raisebox{0.5ex}{\tiny$\substack{+50 \\ -20}$}} & 8  {\raisebox{0.5ex}{\tiny$\substack{+5 \\ -2}$}} \\ 
    88:104 & 500 {\raisebox{0.5ex}{\tiny$\substack{+200 \\ -100}$}} & 1.7  {\raisebox{0.5ex}{\tiny$\substack{+3.3 \\ -1.2}$}} & 24  {\raisebox{0.5ex}{\tiny$\substack{+6 \\ 4}$}} & 36  {\raisebox{0.5ex}{\tiny$\substack{+8 \\ -7}$}} & 5.7  {\raisebox{0.5ex}{\tiny$\substack{+1.2 \\ -1.0}$}} \\ 
    104:116 & 350 {\raisebox{0.5ex}{\tiny$\substack{+100 \\ -60}$}} & 1.9  {\raisebox{0.5ex}{\tiny$\substack{+3.2 \\ -1.3}$}} & 14  {\raisebox{0.5ex}{\tiny$\substack{+3 \\ 3}$}} & 20  {\raisebox{0.5ex}{\tiny$\substack{+5 \\ -4}$}} & 2.5  {\raisebox{0.5ex}{\tiny$\substack{+0.5 \\ -0.5}$}} \\ 
    116:128 & 300 {\raisebox{0.5ex}{\tiny$\substack{+90 \\ -30}$}} & 4.97  {\raisebox{0.5ex}{\tiny$\substack{+0.03 \\ -3.93}$}} & 20  {\raisebox{0.5ex}{\tiny$\substack{+2 \\ 2}$}} & 14  {\raisebox{0.5ex}{\tiny$\substack{+3 \\ -3}$}} & 1.7  {\raisebox{0.5ex}{\tiny$\substack{+0.4 \\ -0.4}$}} \\ 
    \hline
    \end{tabular}
    \tablefoot{The fixed interval, time-resolved fluence is from fitting the spectrum with a COMPT  over a combined (NaI and BGO detectors) spectral range 8\,keV--40\,MeV for \fgbm data and a range 20\,keV--1.2\,MeV for the WIND/KONUS data. The $L_{\rm{iso}}$ and $E_{\rm{iso}}$ values were calculated over the standardized bolometric energy range of 1\,keV to 10\,MeV. The values for GRB\,200415A and GRB\,180128A are consistent with those found in \cite{2021Natur.589..207R} and \cite{Trigg2024A&A...687A.173T}, respectively.}
    \end{table*}  
\FloatBarrier
\twocolumn

\onecolumn
\section{Time-resolved analyses of GRB~200415A and GRB~180128A} \label{time_res_Old_MGFs}
{\renewcommand{\arraystretch}{1.3}
    \begin{table*}[ht!]
    \caption{Time-resolved spectral analysis using BBs.}
    \label{tab:GBM_spec_other}
    \centering
    \begin{tabular}{cccccc}
    \hline\hline
    Time & $E_{\mathrm{p}}$ & $\alpha$ & Energy Flux ($\cal{F}$) & $L_{\mathrm{iso}}$ &  $E_{\mathrm{iso}}$\\
    (ms) & (keV) &  & ($\times10^{-6}$~ergs~s$^{-1}$~cm$^{-2}$) &($\times10^{45}$~erg$\cdot$~s$^{-1}$) &  ($\times10^{44}$~erg)\\ 
    \hline
    \multicolumn{6}{c}{GRB\,200415A}\\
     -4.3:-3.9 & 570 {\raisebox{0.5ex}{\tiny$\substack{+90 \\ -800}$}} & -0.2  {\raisebox{0.5ex}{\tiny$\substack{+0.3 \\ -0.3}$}} & 134  {\raisebox{0.5ex}{\tiny$\substack{+6 \\ -6}$}} & 196  {\raisebox{0.5ex}{\tiny$\substack{+5 \\ -4}$}} & 0.81  {\raisebox{0.5ex}{\tiny$\substack{+0.02 \\ -0.02}$}} \\ 
     -3.9:-3.4 & 800 {\raisebox{0.5ex}{\tiny$\substack{+200 \\ -200}$}} & -0.5  {\raisebox{0.5ex}{\tiny$\substack{+0.3 \\ -0.2}$}} & 230  {\raisebox{0.5ex}{\tiny$\substack{+20 \\ -20}$}} & 341  {\raisebox{0.5ex}{\tiny$\substack{+20 \\ -12}$}} & 1.79  {\raisebox{0.5ex}{\tiny$\substack{+0.01 \\ -0.01}$}} \\ 
     -3.4:-2.9 & 860 {\raisebox{0.5ex}{\tiny$\substack{+120 \\ -100}$}} & -0.33  {\raisebox{0.5ex}{\tiny$\substack{+0.15 \\ -0.15}$}} & 410  {\raisebox{0.5ex}{\tiny$\substack{+20 \\ -20}$}} & 600  {\raisebox{0.5ex}{\tiny$\substack{+20 \\ -10}$}} & 3.1  {\raisebox{0.5ex}{\tiny$\substack{+0.1 \\ -0.1}$}} \\ 
     -2.9:-2.5 & 890 {\raisebox{0.5ex}{\tiny$\substack{+140 \\ -120}$}} & -0.2  {\raisebox{0.5ex}{\tiny$\substack{+0.2 \\ -0.2}$}} & 610  {\raisebox{0.5ex}{\tiny$\substack{+30 \\ -30}$}} & 890  {\raisebox{0.5ex}{\tiny$\substack{+30 \\ -20}$}} & 3.65  {\raisebox{0.5ex}{\tiny$\substack{+0.13 \\ -0.07}$}} \\ 
     -2.5:-0.5 & 1350 {\raisebox{0.5ex}{\tiny$\substack{+150 \\ -130}$}} & -0.3  {\raisebox{0.5ex}{\tiny$\substack{+0.1 \\ -0.1}$}} & 710  {\raisebox{0.5ex}{\tiny$\substack{+20 \\ -30}$}} & 1040  {\raisebox{0.5ex}{\tiny$\substack{+20 \\ -20}$}} & 21.2  {\raisebox{0.5ex}{\tiny$\substack{+0.5 \\ -0.4}$}} \\ 
     -0.5:3.0 & 1800 {\raisebox{0.5ex}{\tiny$\substack{+200 \\ -200}$}} & -0.08  {\raisebox{0.5ex}{\tiny$\substack{+0.08 \\ -0.08}$}} & 610  {\raisebox{0.5ex}{\tiny$\substack{+20 \\ -20}$}} & 900  {\raisebox{0.5ex}{\tiny$\substack{+30 \\ -10}$}} & 32.1  {\raisebox{0.5ex}{\tiny$\substack{+0.7 \\ -0.5}$}} \\ 
     3.0:5.0 & 900 {\raisebox{0.5ex}{\tiny$\substack{+200 \\ -200}$}} & 0.1  {\raisebox{0.5ex}{\tiny$\substack{+0.3 \\ -0.3}$}} & 112  {\raisebox{0.5ex}{\tiny$\substack{+7 \\ -7}$}} & 160 {\raisebox{0.5ex}{\tiny$\substack{+10 \\ -10}$}} & 3.44  {\raisebox{0.5ex}{\tiny$\substack{+0.13 \\ -0.11}$}} \\ 
     5.0:6.5 & 800 {\raisebox{0.5ex}{\tiny$\substack{+300 \\ -300}$}} & 1.0  {\raisebox{0.5ex}{\tiny$\substack{+0.7 \\ -1.1}$}} & 17  {\raisebox{0.5ex}{\tiny$\substack{+5 \\ -6}$}} & 24  {\raisebox{0.5ex}{\tiny$\substack{+5 \\ -4}$}} & 0.4  {\raisebox{0.5ex}{\tiny$\substack{+0.1 \\ -0.1}$}} \\ 
     6.5:22.5 & 1040 {\raisebox{0.5ex}{\tiny$\substack{+50 \\ -50}$}} & 0.8  {\raisebox{0.5ex}{\tiny$\substack{+0.1 \\ -0.1}$}} & 119  {\raisebox{0.5ex}{\tiny$\substack{+5 \\ -5}$}} & 174  {\raisebox{0.5ex}{\tiny$\substack{+5 \\ -3}$}} & 28.5  {\raisebox{0.5ex}{\tiny$\substack{+0.8 \\ -0.5}$}} \\ 
     22.5:65.8 & 800 {\raisebox{0.5ex}{\tiny$\substack{+40 \\ -40}$}} & 0.5  {\raisebox{0.5ex}{\tiny$\substack{+0.1 \\ -0.1}$}} & 50  {\raisebox{0.5ex}{\tiny$\substack{+2 \\ -2}$}} & 73  {\raisebox{0.5ex}{\tiny$\substack{+2 \\ -1}$}} & 32.3  {\raisebox{0.5ex}{\tiny$\substack{+0.6 \\ -0.4}$}} \\ 
     65.8:93.3 & 570 {\raisebox{0.5ex}{\tiny$\substack{+50 \\ -50}$}} & 0.3  {\raisebox{0.5ex}{\tiny$\substack{+0.3 \\ -0.3}$}} & 18.4  {\raisebox{0.5ex}{\tiny$\substack{+1.1 \\ -1.0}$}} & 27  {\raisebox{0.5ex}{\tiny$\substack{+1 \\ -1}$}} & 7.7  {\raisebox{0.5ex}{\tiny$\substack{+0.4 \\ -0.2}$}} \\ 
     93.3:121.2 & 420 {\raisebox{0.5ex}{\tiny$\substack{+40 \\ -30}$}} & 0.8  {\raisebox{0.5ex}{\tiny$\substack{+0.6 \\ -0.5}$}} & 9.2  {\raisebox{0.5ex}{\tiny$\substack{+0.8 \\ -0.8}$}} & 14  {\raisebox{0.5ex}{\tiny$\substack{+1 \\ -1}$}} & 4.0  {\raisebox{0.5ex}{\tiny$\substack{+0.3 \\ -0.2}$}} \\ 
     121.2:150.2 & 250 {\raisebox{0.5ex}{\tiny$\substack{+30 \\ -20}$}} & 1.0  {\raisebox{0.5ex}{\tiny$\substack{+0.7 \\ -0.5}$}} & 3.2  {\raisebox{0.5ex}{\tiny$\substack{+0.8 \\ -0.8}$}} & 5  {\raisebox{0.5ex}{\tiny$\substack{+1 \\ -1}$}} & 1.6  {\raisebox{0.5ex}{\tiny$\substack{+0.3 \\ -0.1}$}} \\ 
    \hline\hline
    \multicolumn{6}{c}{GRB\,180128A}\\
    -12.0:-10.0 & 180 {\raisebox{0.5ex}{\tiny$\substack{+20 \\ -20}$}} & 4.3  {\raisebox{0.5ex}{\tiny$\substack{+1.1 \\ -0.9}$}} & 7  {\raisebox{0.5ex}{\tiny$\substack{+2 \\ -2}$}} & 10  {\raisebox{0.5ex}{\tiny$\substack{+3 \\ -2}$}} & 0.25  {\raisebox{0.5ex}{\tiny$\substack{+0.05 \\ -0.03}$}} \\ 
    -10.0:-7.0 & 510 {\raisebox{0.5ex}{\tiny$\substack{+120 \\ -100}$}} & 0.05  {\raisebox{0.5ex}{\tiny$\substack{+0.58 \\ -0.47}$}} & 37  {\raisebox{0.5ex}{\tiny$\substack{+2 \\ -2}$}} & 54  {\raisebox{0.5ex}{\tiny$\substack{+3 \\ -2}$}} & 1.7  {\raisebox{0.5ex}{\tiny$\substack{+0.1 \\ -0.1}$}} \\ 
    -7.0:-3.0 & 400 {\raisebox{0.5ex}{\tiny$\substack{+40 \\ -60}$}} & 5  {\raisebox{0.5ex}{\tiny$\substack{+2 \\ -3}$}} & 7  {\raisebox{0.5ex}{\tiny$\substack{+2 \\ -3}$}} & 10  {\raisebox{0.5ex}{\tiny$\substack{+2 \\ -1}$}} & 0.5  {\raisebox{0.5ex}{\tiny$\substack{+0.1 \\ -0.1}$}} \\ 
    -3.0:-1.0 & 400 {\raisebox{0.5ex}{\tiny$\substack{+110 \\ -100}$}} & 0.4  {\raisebox{0.5ex}{\tiny$\substack{+1.0 \\ -0.7}$}} & 23  {\raisebox{0.5ex}{\tiny$\substack{+4 \\ -3}$}} & 33  {\raisebox{0.5ex}{\tiny$\substack{+4 \\ -2}$}} & 0.73  {\raisebox{0.5ex}{\tiny$\substack{+0.08 \\ -0.04}$}} \\ 
    -1.0:18.0 & 180 {\raisebox{0.5ex}{\tiny$\substack{+10 \\ -11}$}} & 4.0  {\raisebox{0.5ex}{\tiny$\substack{+0.7 \\ -0.6}$}} & 2.4  {\raisebox{0.5ex}{\tiny$\substack{+0.4 \\ -0.3}$}} & 3.6  {\raisebox{0.5ex}{\tiny$\substack{+0.4 \\ -0.2}$}} & 0.8  {\raisebox{0.5ex}{\tiny$\substack{+0.1 \\ -0.1}$}} \\ 
    18.0:143.0 & 140 {\raisebox{0.5ex}{\tiny$\substack{+20 \\ -20}$}} & 0.8  {\raisebox{0.5ex}{\tiny$\substack{+0.4 \\ -0.3}$}} & 0.5  {\raisebox{0.5ex}{\tiny$\substack{+0.1 \\ -0.1}$}} & 0.7  {\raisebox{0.5ex}{\tiny$\substack{+0.2 \\ -0.1}$}} & 1.0  {\raisebox{0.5ex}{\tiny$\substack{+0.2 \\ -0.1}$}} \\ 
    \hline\hline
    \end{tabular}
    \tablefoot{The BB interval, time-resolved fluence is from fitting the spectrum with a COMPT  over a combined (NaI and BGO detectors) spectral range of 8 \,keV--40\,MeV for \fgbm data. The $L_{\rm{iso}}$ and $E_{\rm{iso}}$ values were calculated over the standardized bolometric energy range of 1\,keV to 10\,MeV. These values are consistent with those found in \cite{2021Natur.589..207R} and \cite{Trigg2024A&A...687A.173T}.}
    \end{table*}}
\FloatBarrier
\twocolumn

\onecolumn
\begin{table*}[h!]
\section{Fit statistics} \label{fit_stats}
    \centering
    \caption{Best-fit model fits for GRB\,231115A.}
    \label{tab:pstat}
    \begin{tabular}{ccccccc}
    \hline\hline
    COMPT & BAND & PL & $\Delta$pstat & $\Delta$pstat & Preferred Model\\
    (pstat/DoF) & (pstat/DoF) & (pstat/DoF) & (PL vs. COMPT) &  (COMPT vs. BAND) & \\ 
    \hline
     \multicolumn{6}{c}{Time-integrated Analysis}\\
     365.869/689.000 & 371.382/688.000 & 437.831/690.000 & 71.962 & 5.513 & COMPT \\
    \hline\hline
    \multicolumn{6}{c}{Time-resolved analysis: Equal time intervals}\\
    241.299/689 & 244.615/688 & 267.104/690 & 25.805 & 3.316 & COMPT \\ 
    252.106/689 & 253.408/688 & 266.932/690 & 14.826 & 1.302 & COMPT \\ 
    221.375/689 & 223.932/688 & 243.074/690 & 21.699 & 2.557 & COMPT \\ 
    171.796/689 & 173.346/688 & 184.005/690 & 12.209 & 1.55 & COMPT \\ 
    177.102/689 & 177.234/688 & 186.445/690 & 9.343 & 0.132 & COMPT \\ 
    227.961/689 & 227.922/688 & 232.894/690 & 4.933 & 0.039 & PL \\ 
    \hline\hline
    \multicolumn{6}{c}{Time-resolved analysis: BB time intervals}\\
    195.191/689 & 195.196/688 & 197.017/690 & 1.826 & 0.005 & PL \\ 
    247.899/689 & 251.096/688 & 277.252/690 & 29.353 & 3.197 & COMPT \\ 
    123.850/689 & 124.486/688 & 129.507/690 & 5.657 & 0.636 & PL \\ 
    214.046/689 & 214.318/688 & 223.112/690 & 9.066 & 0.272 & COMPT \\ 
    300.066/689 & 304.348/688 & 342.067/690 & 42.001 & 4.282 & COMPT \\ 
    208.032/689 & 207.678/688 & 211.964/690 & 3.932 & 0.354 & PL \\ 
    230.227/689 & 230.157/688 & 231.051/690 & 0.824 & 0.07 & PL \\ 
    \hline\hline
    \end{tabular}
    \end{table*}

\section{Optical and infrared follow-up observations} \label{optical_log}

\begin{table*}[ht!]
    \caption{Optical and infrared observations of GRB~231115A.}
    \label{tab:follow-up_optical}
    \centering
    \begin{tabular}{lccccc}
    \hline\hline
    Start Time (UT) & $T-T_0$ (d) & Telescope & Filter & AB Magnitude & Reference  \\
    \hline
    
    2023-11-15 16:36:59 & 0.05 & GOT & $r$ & $>20.1$ &  \citet{GCN35056}  \\
    2023-11-15 16:46:00 & 0.05 & Lulin & $r$ & $>19.2$ &  \citet{GCN35052}  \\  
    2023-11-15 16:47:58 & 0.05 & GIT & $r$ & $>19.3$ &  \citet{GCN35055}  \\ 
    2023-11-15 17:24:23 & 0.075 & MITSuME & $R$ & $>19.5$ &   \citet{GCN35057} \\ 
    2023-11-15 17:37:53 & 0.08 & GRANDMA &  $r$& $>19.3$ &   \citet{GCN35051} \\ 
    2023-11-15 18:16:33 & 0.11  & MITSuME & $R$ & $>20.2$ &   \citet{GCN35057} \\ 
    2023-11-15 22:57:23 & 0.31 & OHP & $R$ & $>21.1$ & \citet{GCN35078}  \\ 
    2023-11-16 00:39:00 & 0.375 & Wendelstein & $r$ & $>22.5$ & This work\\ 
    2023-11-16 00:39:00 & 0.375 & Wendelstein & $i$ & $>22.3$ & This work \\ 
    2023-11-16 00:39:00 & 0.375 & Wendelstein & $J$ & $>19.7$ & This work \\ 
    2023-11-16 03:27:28 & 0.49 & TNG & $r$ & $>22.0$ &   \citet{GCN35077} \\ 
    2023-11-16 03:45:00 & 0.51 & Liverpool & $r$ & $>21.6$ & \citet{GCN35067} \\
    2023-11-22 00:16:32 & 6.36 & Wendelstein & $r$ & -- & This work \\ 
    2023-11-22 00:16:32 & 6.36 & Wendelstein & $i$ & -- & This work \\ 
    2023-11-22 00:16:32 & 6.36 & Wendelstein & $J$ & -- & This work \\ 
    2023-11-22 02:01:48 & 6.43 & Wendelstein & $g$ & $>22.5$ & This work \\
    
    \hline\hline
    \end{tabular}
    \end{table*}
\FloatBarrier
\twocolumn

\onecolumn
\section{Time-resolved spectral fits} \label{sec:time_res_spec_fits}

\begin{figure*}[ht!]
    \begin{minipage}{.5\textwidth}
        \begin{center}
        \includegraphics[width=\linewidth]{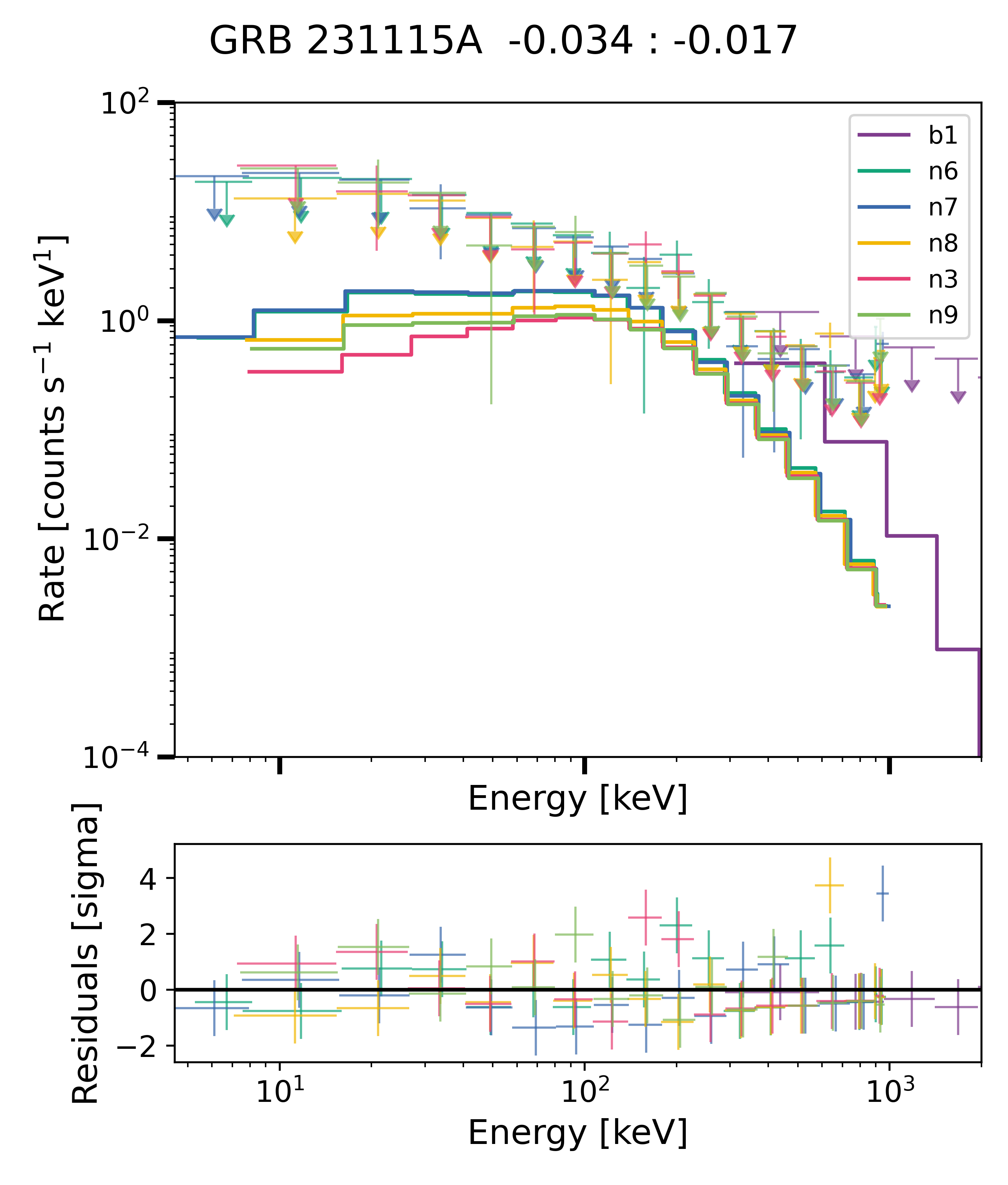}
        \end{center}
    \end{minipage}
    \begin{minipage}{.5\textwidth}
        \begin{center}
        \includegraphics[width=\linewidth]{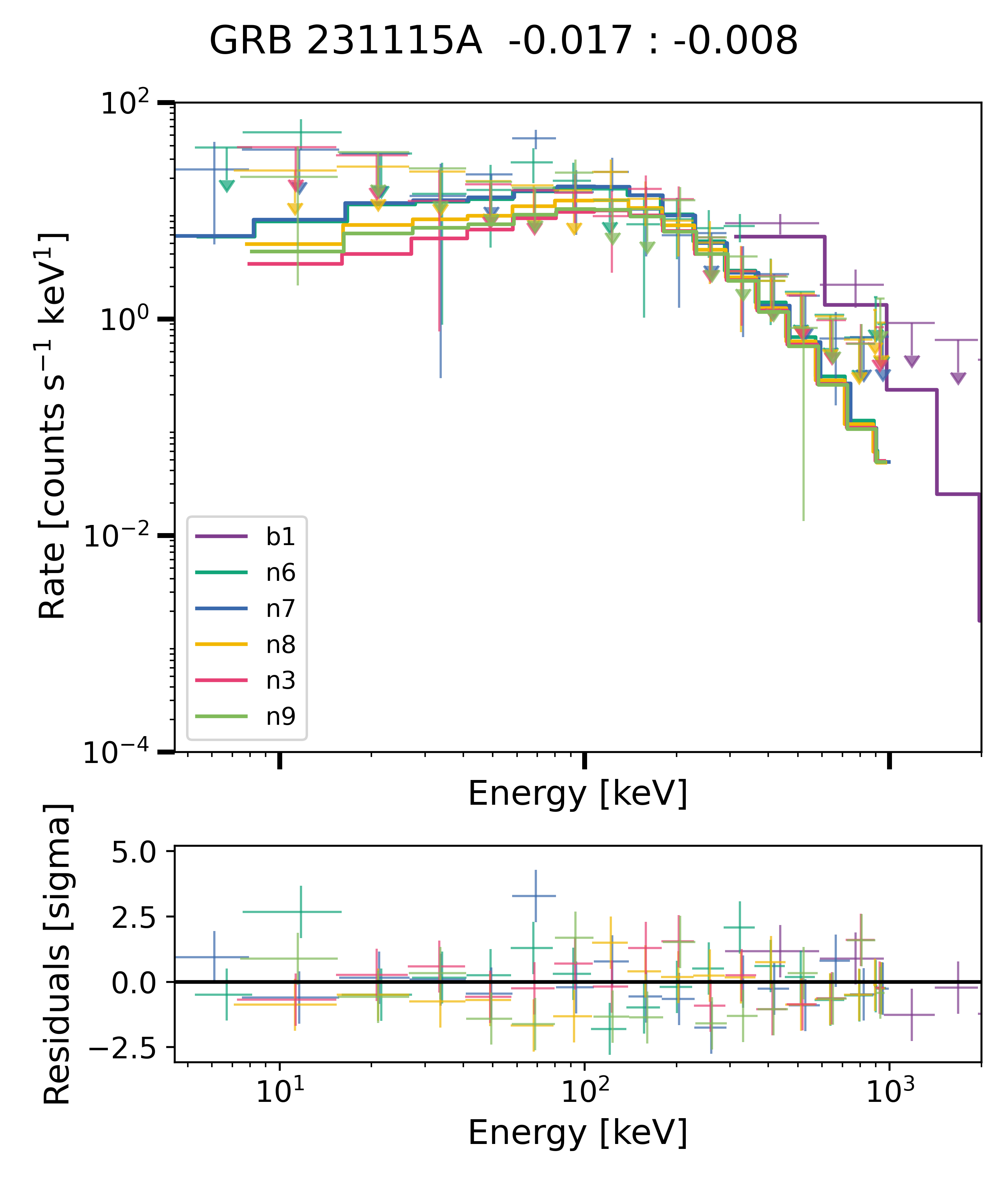}
        \end{center}
    \end{minipage}
    \begin{minipage}{.5\textwidth}
            \begin{center}
            \includegraphics[width=\linewidth]{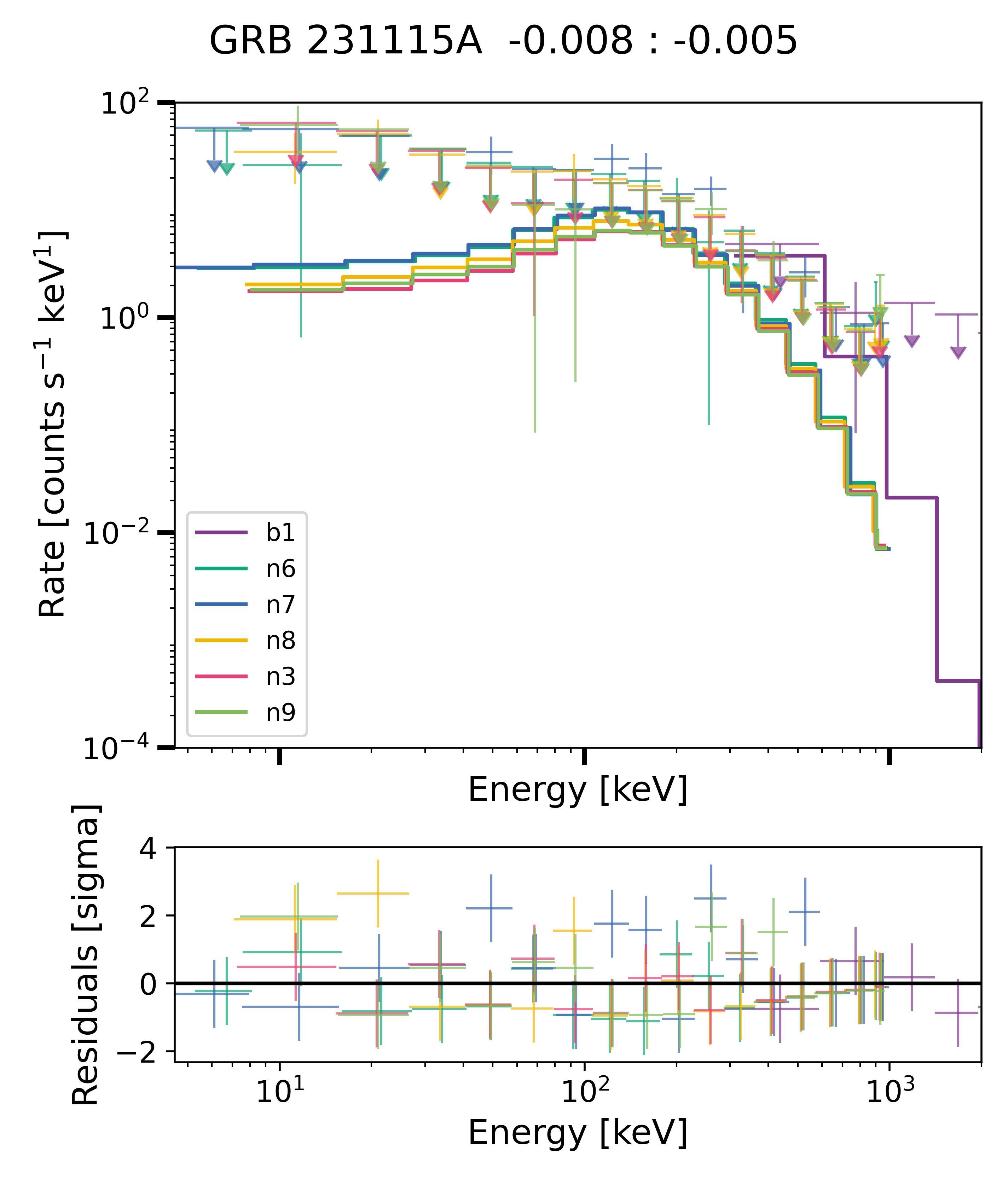}
        \end{center}
    \end{minipage}
    \begin{minipage}{.5\textwidth}
            \begin{center}
            \includegraphics[width=\linewidth]{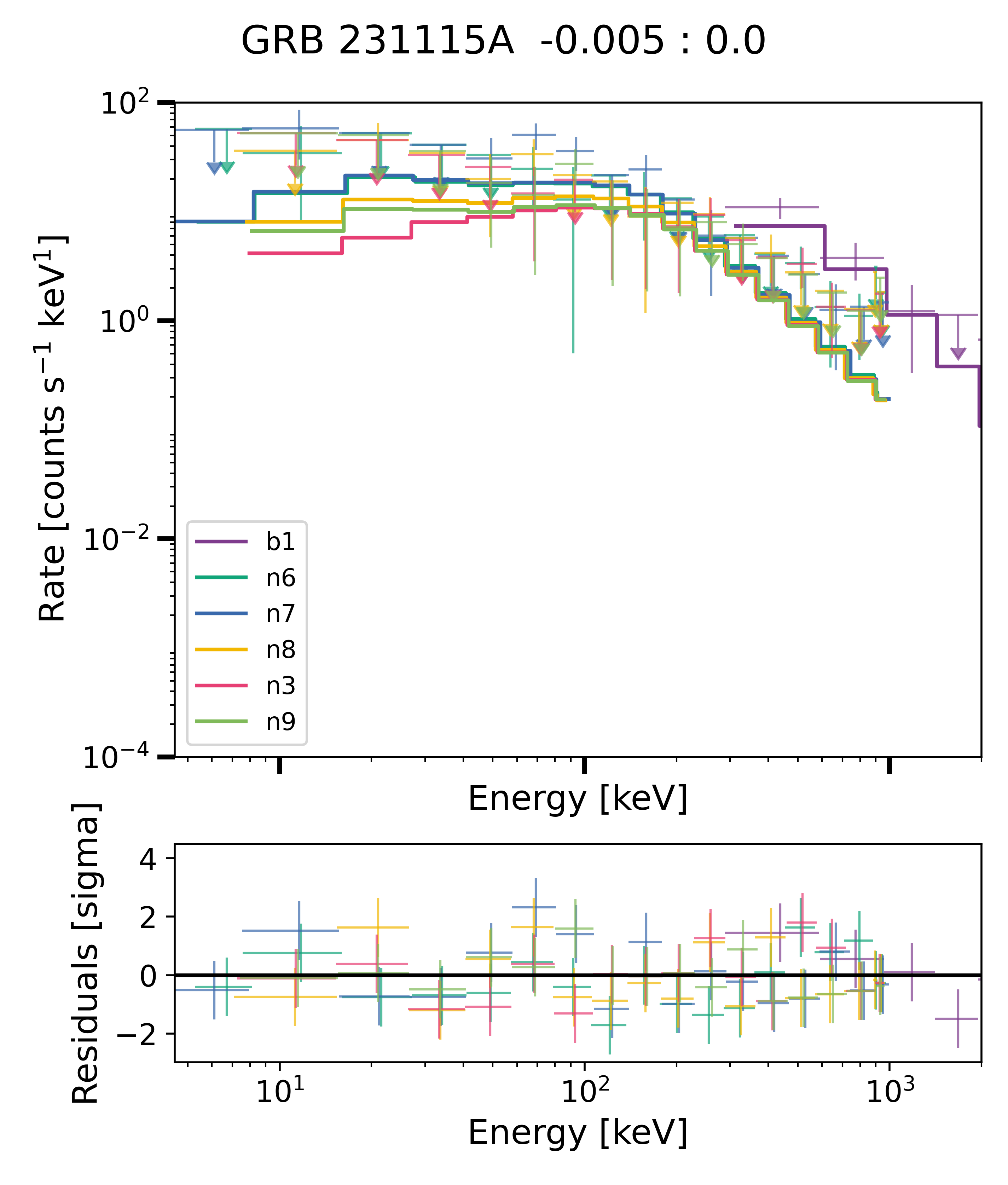}
        \end{center}
    \end{minipage}
    \caption{Time-resolved fits to a COMPT model using the BB time intervals.}
    \end{figure*}
    \begin{figure*}
    \ContinuedFloat
    \begin{minipage}{.5\textwidth}
            \begin{center}
            \includegraphics[width=\linewidth]{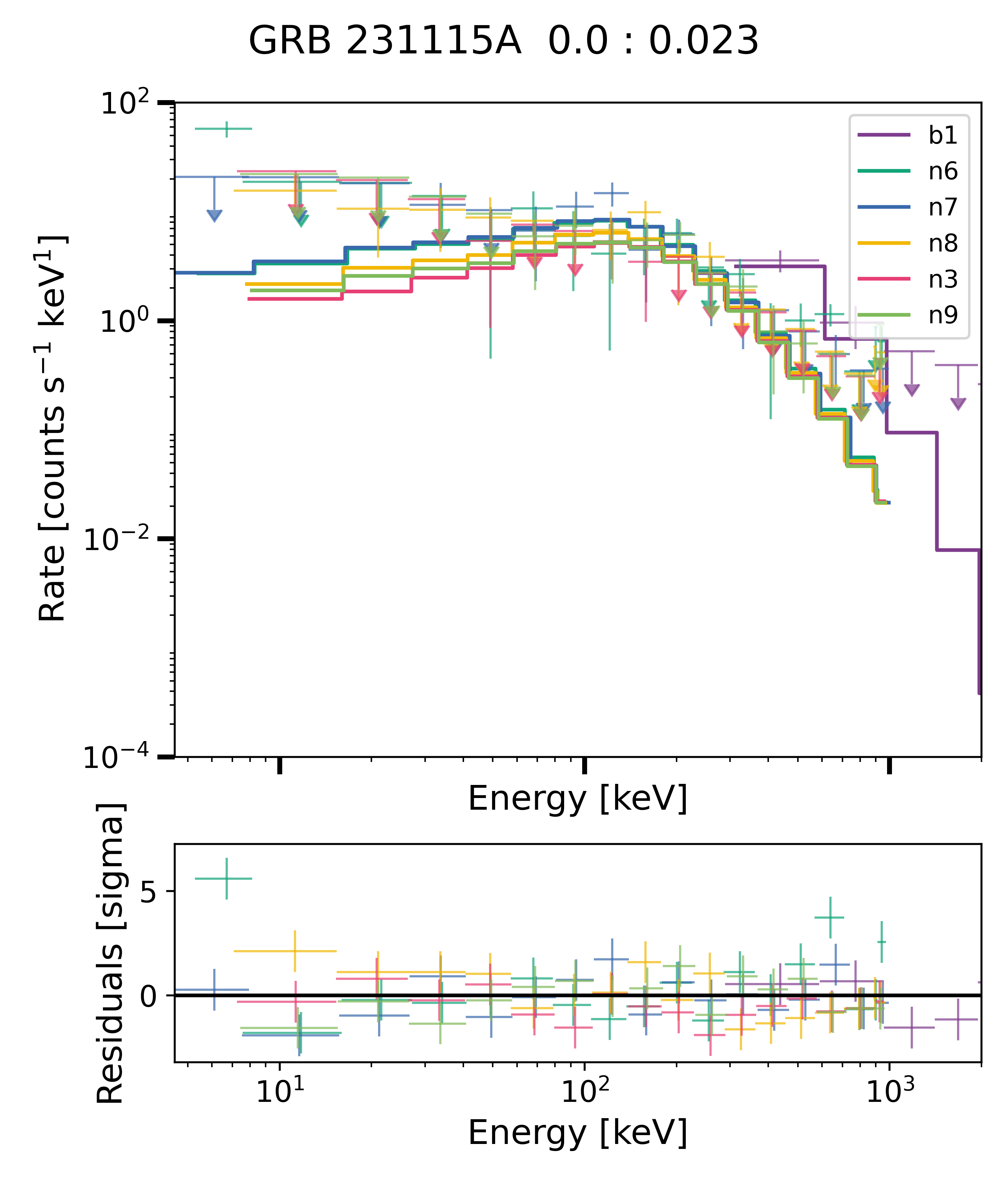}
        \end{center}
    \end{minipage}
    \begin{minipage}{.5\textwidth}
            \begin{center}
            \includegraphics[width=\linewidth]{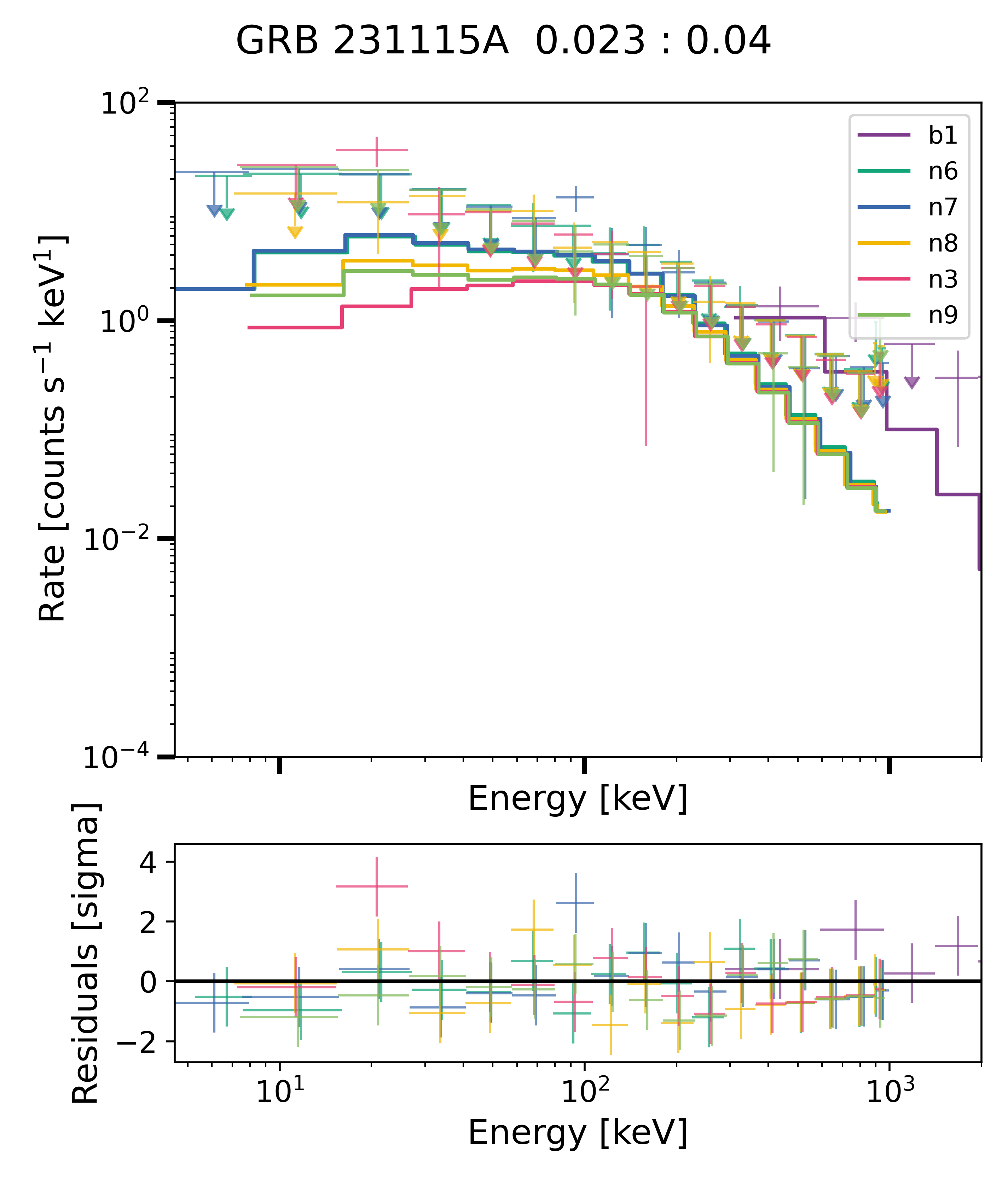}
        \end{center}
    \end{minipage}
    \begin{minipage}{.5\textwidth}
            \begin{center}
            \includegraphics[width=\linewidth]{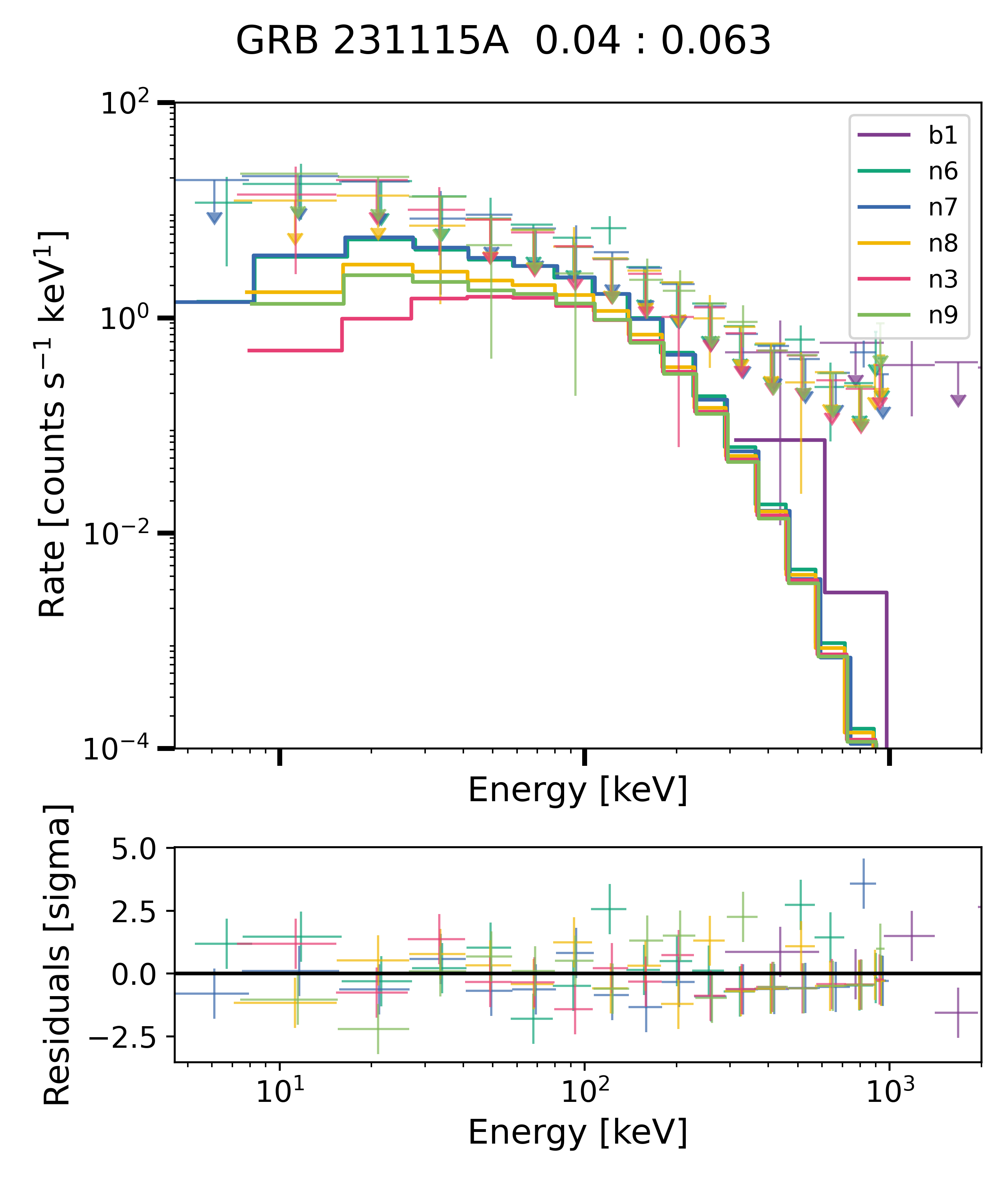}
        \end{center}
    \end{minipage}
    \caption{Continued.}
    \label{fig:time-res_cont}
\end{figure*}

\end{appendix}

\end{document}